\documentclass[fleqn,usenatbib]{mnras}

\DeclareRobustCommand{\VAN}[3]{#2}
\let\VANthebibliography\thebibliography
\def\thebibliography{\DeclareRobustCommand{\VAN}[3]{##3}\VANthebibliography}

\usepackage{graphicx}	% Including figure files
\usepackage{amsmath}	% Advanced maths commands
\usepackage{xspace}

\newcommand{\Websky}{{Websky}\xspace}
\title[CIB and galaxy cluster SZ detection]{Mitigating the impact of the CIB on galaxy cluster SZ detection with spectrally constrained matched filters}

\author[\'{I}. Zubeldia et al.]{
\'{I}\~{n}igo Zubeldia,$^{1,2,3}$\thanks{E-mail: inigo.zubeldia@ast.cam.ac.uk}
Jens Chluba,$^{1}$ and Richard Battye$^{1}$
\\
$^{1}$Jodrell Bank Centre for Astrophysics, University of Manchester, Manchester M13 9PL UK\\
$^{2}$Institute of Astronomy, University of Cambridge, Madingley Road, Cambridge CB3 0HA\\
$^{3}$Kavli Institute for Cosmology, University of Cambridge, Madingley Road, Cambridge CB3 0HA
}

\date{Accepted XXX. Received YYY; in original form ZZZ}

\pubyear{2022}

\begin{document}
\label{firstpage}
\pagerange{\pageref{firstpage}--\pageref{lastpage}}
\maketitle

% Abstract of the paper
\begin{abstract}
Galaxy clusters detected through the thermal Sunyaev-Zeldovich (tSZ) effect are a powerful cosmological probe from which constraints on cosmological parameters such as $\Omega_{\mathrm{m}}$ and $\sigma_8$ can be derived. The measured cluster tSZ signal can be, however, contaminated by Cosmic Infrared Background (CIB) emission, as the CIB is spatially correlated with the cluster tSZ field. We quantify the extent of this contamination by applying the iterative multi-frequency matched filter (iMMF) cluster-finding method to mock \textit{Planck}-like data from the \Websky simulation. We find a significant bias in the retrieved cluster tSZ observables (signal-to-noise and Compton-$y$ amplitude), at the level of about $0.5\, \sigma$ per cluster. This CIB-induced bias translates into about $20$\% fewer detections than expected if all the \textit{Planck} HFI channels are used in the analysis, which can potentially bias derived cosmological constraints. We introduce a spectrally constrained iMMF, or sciMMF, which proves to be highly effective at suppressing this CIB-induced bias from the tSZ cluster observables by spectrally deprojecting the cluster-correlated CIB at the expense of a small signal-to-noise penalty. Our sciMMF is also robust to modelling uncertainties, namely to the choice of deprojection spectral energy distribution. With it, CIB-free cluster catalogues can be constructed and used for cosmological inference. We provide a publicly available implementation of our sciMMF as part of the \texttt{SZiFi} package.
\end{abstract}

% Select between one and six entries from the list of approved keywords.
% Don't make up new ones.

%%%%%%%%%%%%%%%%%%%%%%%%%%%%%%%%%%%%%%%%%%%%%%%%%%

%%%%%%%%%%%%%%%%% BODY OF PAPER %%%%%%%%%%%%%%%%%%

\section{Introduction}

%SZ detected clusters have become probe

Galaxy clusters detected through the thermal Sunyaev-Zeldovich (tSZ) effect (\citealt{Sunyaev1972}; see \citealt{Carlstrom2002} and \citealt{Mroczkowski2019} for reviews) are a powerful cosmological probe from which constraints on cosmological parameters such as $\Omega_{\mathrm{m}}$ and $\sigma_8$ can be derived. Over the past decade, a number of galaxy cluster catalogues with $\sim10^2$--$10^3$ objects have been constructed from \textit{Planck}, ACT, and SPT data (e.g., \citealt{Vanderlinde2010,Hasselfield2013,Planck2014,Bleem2015,Planck2016xxvii,Tarrio2019,Bleem2020,Aghanim2019,Hilton2020, Melin2021}) and subsequently used in cosmological number count analyses (e.g., \citealt{Vanderlinde2010,Hasselfield2013,Planck2014XX,Bleem2015,Ade2016,Bocquet2018,Zubeldia2019,Salvati2021,Chaubal2022}). In coming years, experiments such as the Simons Observatory (SO, \citealt{SO2019}) and CMB-S4 \citep{Abazajian2016} are expected to deliver over an order of magnitude more objects than their predecessors. These upcoming samples will have an unprecedented potential for constraining cosmological parameters, but for it to be realised, systematics across the full analysis pipeline will have to be understood to a much higher level of accuracy than is presently the case. %In particular, accurately modelling of the cluster observable(s),  to the cluster mass, which is what is predicted by the theory through the halo mass function, will be crucial for the cosmological success of these upcoming cluster samples \citep{Pratt2019}. 
%Dust emission (CIB) may be a problem. In particular, MMFs.

%Thus emitted by the clusters' galaxies, by galaxies in neighbouring regions, or by the intra-cluster medium. 

A potential systematic affecting the tSZ cluster detection and characterisation process is dust emission that is spatially correlated with the detected clusters, which can be thought of as the contribution to the Cosmic Infrared Background (CIB; see, e.g., \citealt{Lagache2005}) that is spatially correlated with them. Dust emission at the location of galaxy clusters has indeed been detected using IRAS and \textit{Planck} data in stacking analyses \citep{Montier2005,Giard2008,Planck2016XLIII,Planck2016XXIII,Melin2018}, and dust has also been observed to bias measurements of the tSZ signal of galaxies \citep{Planck2013XI} and of active galactic nuclei \citep{Soergel2017}. In addition, the CIB--tSZ correlation has been measured in power spectrum analyses \citep{Planck2016XXIII} and has been shown to bias maps of the tSZ signal (Compton-$y$ maps) and, especially, their cross-correlation with other tracers of the density field (e.g., \citealt{Yan2019,Bleem2022,Sanchez2022}). The impact of the cluster-correlated CIB on cluster tSZ detection, however, has remained little discussed in the literature, there being only one detailed analysis, that of \citet{Melin2018}. In it, clusters with dust emission from a model calibrated to \textit{Planck} data were injected to \textit{Planck} data and used to assess the impact on the retrieved cluster number counts and cosmological constraints. This was found to be small, with about 10\,\% fewer detections in the number counts between $z=0.4$ and $z=0.8$ relative to the case with no dust contamination and negligible changes in the derived cosmological parameters.

In this work we revisit this issue, the impact of the cluster-correlated CIB emission on tSZ cluster detection, in the context of \textit{Planck} using the state-of-the-art \Websky simulation \citep{Stein2019,Stein2020}. As the cluster detection method we use multi-frequency matched filters (MMFs; \citealt{Herranz2002,Melin2006}), the standard tool with which clusters are blindly detected in SZ surveys. In particular, we use the iterative MMF of \citet{Zubeldia2022} as implemented in \texttt{SZiFi}, the Sunyaev-Zeldovich iterative Finder, and apply it to mock \textit{Planck}-like observations produced with the \Websky sky maps. We find significant biases in the cluster observables, survey completeness, and cluster number counts, with about 20\,\% fewer detections if the six \textit{Planck} HFI channels are used in the analysis relative to the case with no cluster-correlated CIB emission. These biases could potentially result in biased cosmological constraints, as we argue here. We introduce a spectrally constrained iterative MMF, or sciMMF, which is designed to completely null or `deproject' the contributions from a set of foregrounds with given spectral energy distributions (SEDs). We apply it to our mock data, deprojecting the cluster-correlated CIB, and find that it can be highly effective at suppressing the CIB-induced biases at the expense of a small signal-to-noise penalty. Our sciMMF thus emerges as a promising tool with which to construct CIB-free cluster catalogues using data from existing (e.g., \textit{Planck}) and upcoming (e.g., Simons Observatory; \citealt{SO2019}) experiments, from which more accurate cosmological constraints can in turn be obtained.

%mention moment expansion?

This paper is organised as follows. First, in Section \ref{sec:theory} we introduce our spectrally constrained MMF and illustrate its performance in a toy model scenario. Next, we describe our MMF implementation, \texttt{SZiFi}, in Section \ref{sec:implementation}, and our simulated maps and cluster catalogues in Section \ref{sec:sims}. We then analyse our cluster catalogues in Section \ref{sec:results}, showing that the cluster-correlated CIB can bias significantly the cluster observables, the survey completeness, and the cluster number counts (Section \ref{subsec:immf}), and that our sciMMF is highly effective at suppressing these biases (Section \ref{subsec:sciMMF}) at the cost of a small signal-to-noise penalty (Section \ref{subsec:bias_snr}). We also investigate the robustness of our results to the choice of deprojection SED (Section \ref{subsec:robustness}) and the adequacy of our sciMMF given the spatial nature of the cluster-correlated CIB (Section \ref{subsec:pointsource}). Finally, in Section \ref{subsec:forecasts} we present a set of forecasts for the loss of signal-to-noise due to CIB deprojection for ACT, SPTpol, and the Simons Observatory, and we conclude in Section \ref{sec:conclusion}. %Mention appendices?

\section{Spectrally constrained multi-frequency matched filters}\label{sec:theory}

In this section we introduce the spectrally constrained multi-frequency matched filter, deriving it from the constrained internal linear combination (ILC) method (Section \ref{subsec:formalism}). We then illustrate its performance in a toy model scenario (Section \ref{subsec:toymodel}) and discuss how to mitigate the impact of errors in the spectral energy distribution (SED) to be deprojected with the moment expansion approach (Section \ref{subsec:moment}).

\subsection{Derivation}\label{subsec:formalism}

Let us consider a set of beam-deconvolved intensity maps at different frequencies, $\bmath{d}(p)$, where the vector dimension of $\bmath{d}$ is equal to the number of frequency channels $N_{\mathrm{f}}$ and where $p$ denotes pixel $p$, which can be a pixel in real space, harmonic space, etc. We assume that  $\mathbfit{d}(p)$ can be written as
%--------------------------------
\begin{equation}
    \mathbfit{d} (p) = \mathbfit{a}  y (p) +  \mathbfss{A}_{\mathrm{f}} \mathbfit{s}_{\mathrm{f}} (p) +  \mathbfit{n} (p).
\end{equation}
%--------------------------------
Here, $y (p)$ is the signal that is being targeted (in our case, the tSZ Compton-$y$) and $\mathbfit{a}$ is its SED (in our case, the tSZ SED); $\mathbfss{A}_{\mathrm{f}}$ is the foreground mixing matrix, containing the SEDs of all the foregrounds present in the data, $\mathbfss{A}_{\mathrm{f}} = [ \mathbfit{b}_1 \dots \mathbfit{b}_{N_{\mathrm{f}}} ] $, where $\mathbfit{b}_i$ is the SED of the ith foreground; $\mathbfit{s}_{\mathrm{f}} (p)$ is a vector of dimension $N_{\mathrm{f}}$, where $N_{\mathrm{f}}$ is the number of foregrounds, describing the spatial variation of the foregrounds; and $\mathbfit{n} (p)$ is some additive noise.

A set of weights $\mathbfit{w}$ can be constructed such that $\hat{y} =  \mathbfit{w}^T \mathbfit{d}$ is an estimator of $y$. The simplest solution is the `standard' internal linear combination (ILC), in which the variance of the output map, $\sigma_y^2 = \mathbfit{w}^T \mathbfss{C} \mathbfit{w}$, where $\mathbfss{C}$ is the covariance matrix of the data, is minimised subject to the constraint that there is unit response to the component of interest, i.e., $\mathbfit{w}^T \mathbfit{a} = 1$. The ILC weights are given by \citep[e.g.,][]{Tegmark1996,Eriksen2004}
%---------------------
\begin{equation}\label{ilc}
    \mathbfit{w}_{\mathrm{ILC}}^T = ( \mathbfit{a}^T \mathbfss{C}^{-1} \mathbfit{a})^{-1} \mathbfit{a}^T \mathbfss{C}^{-1}.
\end{equation}
%---------------------
By construction, the ILC estimate of $y$ has the lowest variance of all possible estimators of the form $\hat{y} =  \mathbfit{w}^T \mathbfit{d}$, but it can be biased, in particular if the foregrounds are spatially correlated with the component of interest (e.g., \citealt{Remazeilles2011}). 

Alternatively, the contributions from one or several foregrounds can be nulled by imposing the additional constraints $\mathbfit{w}^T \mathbfit{b}_i = 0$ for $i = 0$, \dots, $i=N_{\mathrm{dep}}$, where $N_{\mathrm{dep}}$ is the number of foregrounds to be nulled or `deprojected'. This leads to a constrained ILC (cILC; \citealt{Remazeilles2011}), whose weights are given by (e.g., \citealt{Remazeilles2011,Abylkairov2021})
%---------------------
\begin{equation}\label{cilc}
    \mathbfit{w}_{\mathrm{cILC}}^T = \mathbfit{c}^T (\mathbfss{A}^T \mathbfss{C}^{-1} \mathbfss{A})^{-1} \mathbfss{A}^T \mathbfss{C}^{-1},
\end{equation}
%---------------------
where $\mathbfss{A} = [\mathbfit{a} \,\, \mathbfit{b}_1 \dots \mathbfit{b}_{N_{\mathrm{dep}}} ] $ is a $N_{\mathrm{f}} \times N_{\mathrm{dep}} + 1$ mixing matrix and $\mathbfit{c}^T = [1 \,\, 0 \dots 0]$ is a vector of dimension $N_{\mathrm{dep}} + 1$ that ensures that the correct component is retrieved. While deprojecting one or several foregrounds can be desirable to remove biases, this comes at the expense of an inevitable loss of signal-to-noise, as some of the degrees of freedom that are used for variance minimisation in the standard ILC are now used for foreground removal.

Let us now assume that the Compton-$y$ map due to a galaxy cluster can be written as $y (p) = y_0 y_{\mathrm{t}} (p ;  \theta_{500}, \bmath{\theta}_\mathrm{c} )$, where $y_0$ is an amplitude parameter, which we take to be the value of the cluster's Compton-$y$ parameter at the cluster centre, and $y_{\mathrm{t}} (\mathbfit{l} ; \theta_{500}, \bmath{\theta}_\mathrm{c})$ is a spatial template. This template depends on the cluster angular size $\theta_{500}$, defined, as is customary, as the angle subtended by $R_{500}$, the radius within which the mean density is equal to 500 times the critical density at the cluster's redshift. It also depends on the sky coordinates of the cluster centre, $\bmath{\theta}_\mathrm{c}$. Assuming the flat-sky approximation and the covariance of the data to be isotropic, a matched filter estimator for the amplitude parameter $y_0$ can then be written as
%---------------------
\begin{equation}
    \hat{y}_0 =   \left[ \int \frac{ d^2 \mathbfit{l}}{2 \pi}  \frac{ y_{\mathrm{t}}^{\ast} (\mathbfit{l}) y_{\mathrm{t}} (\mathbfit{l}) }{\sigma_y^2 (l)} \right]^{-1} \int \frac{ d^2 \mathbfit{l}}{2 \pi} \frac{y_{\mathrm{t}}^{\ast} (\mathbfit{l}) \hat{y} (\mathbfit{l}) }{\sigma_y^2(l)},
\end{equation}
%---------------------
where the inverse-variance weighting ensures optimality. Substituting the general ILC expressions for $\hat{y}$ and $\sigma_y^2$, $\hat{y} = \mathbfit{w}^T \mathbfit{d}$ and $\sigma_y^2 = \mathbfit{w}^T \mathbfss{C} \mathbfit{w}$, respectively, leads to the following expression,

\begin{equation}\label{mmf_general}
    \hat{y}_0 = N^{-1} \int \frac{ d^2 \mathbfit{l}}{2 \pi} \frac{y_{\mathrm{t}}^{\ast} (\mathbfit{l}) \mathbfit{w}^T(l) \mathbfit{d} (\mathbfit{l})}{\mathbfit{w}^T  (l) \mathbfss{C}(l) \mathbfit{w} (l)},
\end{equation}
where the normalisation $N$ is given by
%---------------------
\begin{equation}\label{norm}
   N =  \int \frac{ d^2 \mathbfit{l}}{2 \pi} \frac{ y_{\mathrm{t}}^{\ast} (\mathbfit{l}) y_{\mathrm{t}} (\mathbfit{l}) }{\mathbfit{w}^T(l) \mathbfss{C} (l) \mathbfit{w} (l)}.
\end{equation}
%---------------------
This constitutes a general multi-frequency matched filter (MMF) estimator, which can be thought of as operating directly on the experiment multifrequency maps. The MMF estimator variance, $\sigma^2_{y_0}$, is given by $\sigma^2_{y_0} = N^{-1}$, as can be easily shown, and the signal-to-noise of a detection, $q$, by
%---------------------
\begin{equation}\label{snr}
    q = \frac{\hat{y}_0}{\sigma_{y_0}} =  N^{-1/2} \int \frac{ d^2 \mathbfit{l}}{2 \pi} \frac{y_{\mathrm{t}}^{\ast} (\mathbfit{l}) \mathbfit{w}^T(l) \mathbfit{d} (\mathbfit{l})}{\mathbfit{w}^T  (l) \mathbfss{C}(l) \mathbfit{w} (l)}.
\end{equation}
%---------------------
We can also define the true signal-to-noise of a detection, $\bar{q}_{\mathrm{t}}$, as the signal-to-noise that would be measured in the absence of foregrounds and noise, but assuming the full covariance of the data in the MMF, and with the matched filter template evaluated at the true parameter values (i.e., true angular size and sky location). This is equal to the ensemble-averaged signal-to-noise $q$ if the foregrounds and the noise have zero mean (or if the $\mathbfit{l} = 0$ mode is not used in the matched filter) and are uncorrelated with the cluster, and is given by
%---------------------
\begin{equation}\label{snr_true}
    \bar{q}_{\mathrm{t}} = \frac{y_0}{\sigma_{y_0}} =  N^{-1/2} \int \frac{ d^2 \mathbfit{l}}{2 \pi} \frac{y_{\mathrm{t}}^{\ast} (\mathbfit{l}) y (\mathbfit{l})}{\mathbfit{w}^T  (l) \mathbfss{C}(l) \mathbfit{w} (l)}.
\end{equation}
%---------------------
If the noise covariance is well determined, the observed signal-to-noise will have unit standard deviation around $\bar{q}_{\mathrm{t}}$ (see \citealt{Zubeldia2022} for a detailed discussion).

We note that the MMF expressions given in Eqs. (\ref{mmf_general})--(\ref{snr_true}) are also valid if the data $\mathbfit{d} (\mathbfit{l})$ is convolved by the instrument beam, provided that so are the covariance matrix and the spatial template $y_{\mathrm{t}} (\mathbfit{l})$, allowing for a more numerically stable implementation. In this work, we always use beam-convolved MMFs, as implemented in \texttt{SZiFi} (see Section \ref{sec:implementation}).

%, and if the normalisation $N$, given by Eq. (\ref{norm}), is computed with one of the two template factors appearing in the numerator being beam-convolved.

%Specify to standard MMF and constrained MMF

If we now substitute the expression for the standard ILC weights, given in Eq. (\ref{ilc}), into our general MMF formulae, we recover the `standard' MMF \citep{Melin2006}, which has been used widely for galaxy cluster detection with multi-frequency CMB data (e.g., \citealt{Planck2016xxvii,Bleem2020,Hilton2020}). For completeness, the $y_0$ estimator reads
%---------------------
\begin{equation}\label{estimator}
\hat{y}_{0}^{\mathrm{MMF}} =  (N^{\mathrm{MMF}})^{-1} \int  \frac{ d^2 \mathbfit{l}}{2 \pi}  \,  \mathbfit{a} y_{\mathrm{t}}^{\ast}  (\mathbfit{l})  \mathbfss{C}^{-1} (l) \mathbfit{d} (\mathbfit{l}),
\end{equation}
%---------------------
with
%---------------------
\begin{equation}\label{noise}
   N^{\mathrm{MMF}} =   \int  \frac{ d^2 \mathbfit{l}}{2 \pi} \, y_{\mathrm{t}}^{\ast} (\mathbfit{l})  \mathbfit{a}^{\dagger}  \mathbfss{C}^{-1} (l) \mathbfit{a} y_{\mathrm{t}}  (\mathbfit{l})
   .
\end{equation}
%---------------------
%where we recall that the signal SED, $\mathbfit{a} (\mathbfit{l})$, is beam-convolved.
%
On the other hand, if we instead use the constrained ILC weights, given in Eq. (\ref{cilc}), we obtain a \emph{spectrally constrained} MMF, or scMMF. This estimator can be written as
%---------------------
\begin{equation}
\hat{y}_{0}^{\mathrm{scMMF}} =  (N^{\mathrm{scMMF}})^{-1} \int  \frac{ d^2 \mathbfit{l}}{2 \pi}  \, \frac{y_{\mathrm{t}}^{\ast}   \mathbfit{c}^T (\mathbfss{A}^T \mathbfss{C}^{-1} \mathbfss{A})^{-1} \mathbfss{A}^T  \mathbfss{C}^{-1}  \mathbfit{d} }{\mathbfit{c}^T (\mathbfss{A}^T \mathbfss{C}^{-1} \mathbfss{A})^{-1} \mathbfit{c}},
\end{equation}
%---------------------
where
%---------------------
\begin{equation}\label{norm}
   N^{\mathrm{scMMF}} =  \int \frac{ d^2 \mathbfit{l}}{2 \pi} \frac{ y_{\mathrm{t}}^{\ast}  y_{\mathrm{t}}  }{\mathbfit{c}^T (\mathbfss{A}^T \mathbfss{C}^{-1} \mathbfss{A})^{-1} \mathbfit{c}},
\end{equation}
%---------------------
and where we note that we have left the $\mathbfit{l}$-dependence of the different factors in the integrand implicit in order to avoid clutter in the notation. As in the general case, the estimator variance is given by $(\sigma^{\mathrm{scMMF}}_{y_0})^2 = (N^{\mathrm{scMMF}})^{-1}$, and the signal-to-noise by $q_{\mathrm{scMMF}} = \hat{y}_{0}^{\mathrm{scMMF}} / \sigma^{\mathrm{scMMF}}_{y_0}$.

Since it is constructed from a constrained ILC, the scMMF estimator completely nulls the contribution from one or several foregrounds with SEDs of choice. This is done at the expense of a signal-to-noise penalty, which we note is independent of the amplitude of the foreground(s) to be deprojected, but depends on the shape of its (or their) SED(s). We choose to refer to this MMF estimator as a spectrally constrained MMF, rather than just constrained, as other matched filter estimators have been proposed in which a spatial template for the signal to be deprojected is also assumed (in particular, see the constrained matched filters of \citealt{Erler2019}). Our estimator, however, makes no assumptions about the spatial nature of the foregrounds to be nulled, only requiring knowledge of their SEDs. In addition, we note that other MMF estimators can be constructed based on other ILC methods, e.g., the partially-constrained ILC of \citealt{Abylkairov2021}, by simply substituting the appropriate weights $\mathbfit{w}$ into Eq. (\ref{mmf_general}). We leave the investigation of these other possible MMFs to future work.

%This can be useful for deprojecting the cluster-correlated CIB emission, which is the main focus of this work, as its spatial profile is not very well constrained and it cannot be considered a point source for an experiment with the resolution of \textit{Planck} or higher, as discussed in Section \ref{subsec:pointsource}.
%In Appendix \ref{appendix} we explore the relationship between our scMMF and the constrained MMF of \citealt{Erler2019} further, deriving our scMMF by imposing a set of constraints on the variance minmisation problem, rather than from the cILC weights.

%Stress that loss of SNR is independent from the amplitude of the signal to be removed

Finally, we note that in order to construct any of the MMFs discussed here, knowledge of the covariance matrix $\mathbfss{C} (l)$ is required. The covariance is typically estimated from the data and taken to be equal to the cross-channel power spectra of the data (e.g., \citealt{Planck2016xxvii}). However, as discussed extensively in \citealt{Zubeldia2022}, doing so leads to biases and to a loss of signal-to-noise. As shown in \citet{Zubeldia2022}, these effects are eliminated if an iterative approach is followed, in which the clusters detected in a first step are masked out and the noise covariance is re-estimated, leading to an updated, iterative cluster catalogue. In this work we will always follow this iterative approach, and will therefore use a `standard' iterative MMF, or iMMF (introduced in \citealt{Zubeldia2022}), and a spectrally constrained iterative MMF, or sciMMF.

\subsection{Numerical illustration with a toy model}\label{subsec:toymodel}

In order to illustrate the performance of our spectrally constrained MMF, we produce and analyse a set of idealised mock observations. More specifically, we generate maps of single clusters at the six \textit{Planck} High Frequency Instrument (HFI) channels (100--857\,GHz), with a square field of $256 \times 256$ pixels for each cluster. Each field is centred at the cluster centre and we choose a pixel size of 1\,arcmin. We consider a single cluster mass of $M_{500} = 5 \times 10^{14} M_{\odot}$ at redshift $z = 0.2$ and assume the cluster pressure profile of \citet{Arnaud2010}. As a toy foreground we add a Gaussian signal centred at the cluster centre with a FWHM of $10$\,arcmin and the SED of the CIB of the Wesbky simulation at $z=0.2$ (\citealt{Stein2020}; see Section \ref{sec:sims}). This is intended to be a toy model of the cluster-correlated CIB emission at the redshift of a typical \textit{Planck} cluster (see Section \ref{subsec:pointsource}). The amplitude of this Gaussian signal is given by a multiplicative parameter, $A_{\mathrm{CIB}}$. We consider six values of $A_{\mathrm{CIB}}$, logarithmically spaced across 1.5 decades, chosen so that the bias in the standard MMF estimator ranges from a few percent to order unity. At each frequency, the cluster tSZ signal and the CIB-like foreground are convolved by the corresponding isotropic \textit{Planck} beam (available in the Planck Legacy Archive, PLA\footnote{\texttt{pla.esac.esa.int}}). Finally, white noise with \textit{Planck}-like noise levels is added (see Section \ref{sec:sims} for the exact noise levels we assume). We produce 100 such mock observations for each of the six values of $A_{\mathrm{CIB}}$ considered.

For each field, we construct two MMF estimators, a standard MMF and a scMMF. We use the same covariance for the two MMFs, which we estimate from the white noise map directly, binning the cross-frequency power spectra in a set of eight bins linearly spaced between $l=0$ and $l=2700$. Using this approach we exclude the cluster tSZ signal from the covariance, thus effectively making the two MMFs iterative. For both MMFs we take as the spatial template the input Compton-$y$ map, and for the scMMF we take the input CIB SED as the SED to be deprojected. In addition, we evaluate both MMFs at the true cluster location, i.e., at the centre of each map. With this set-up, we produce two sets of 100 $\hat{y}_0$ and signal-to-noise $q$ measurements for each value of the CIB amplitude $A_{\mathrm{CIB}}$.

\begin{figure}
\centering
\includegraphics[width=0.4\textwidth,trim={00mm 5mm 0mm 15mm},clip]{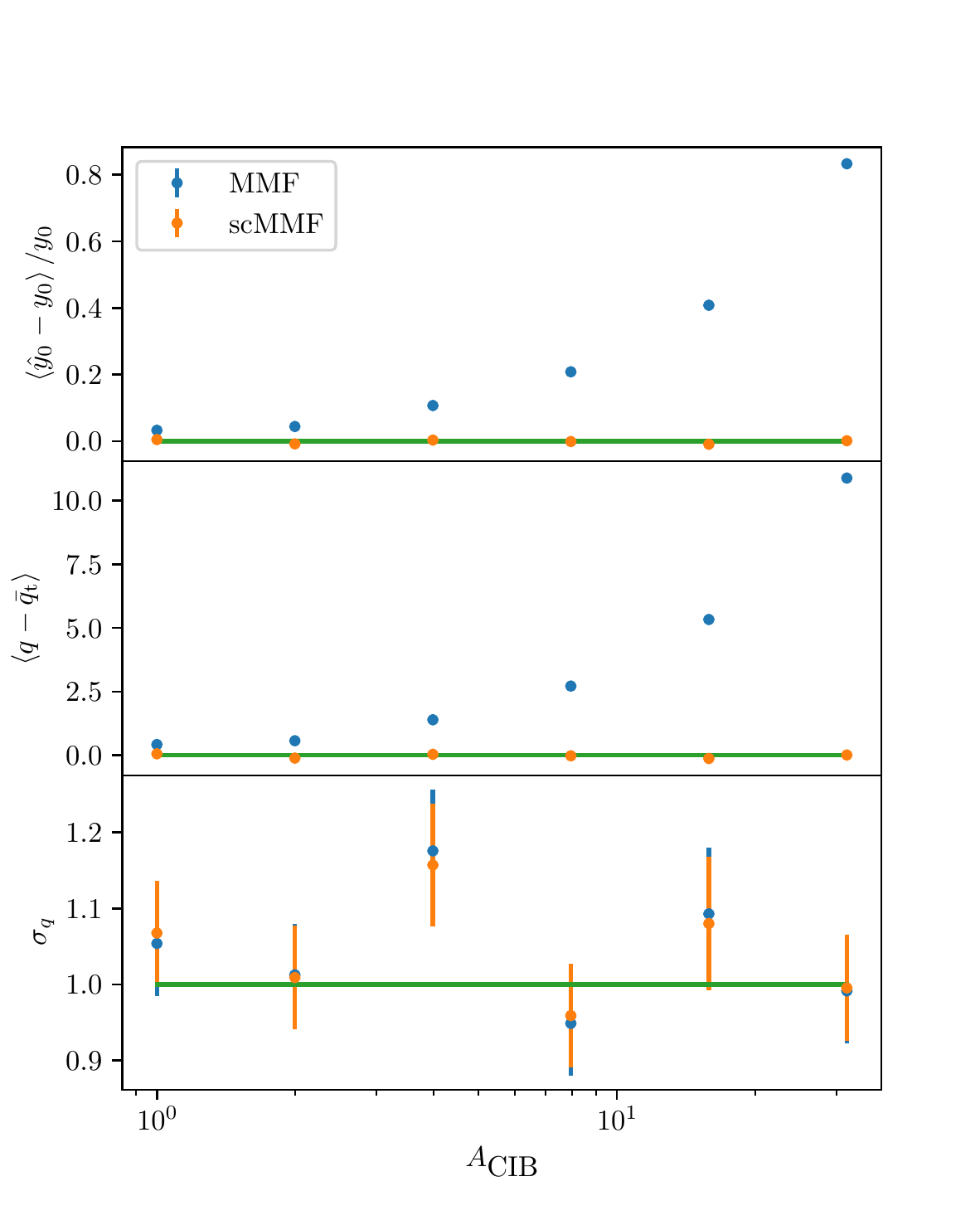}
\caption{Comparison of the performance of the standard MMF and the scMMF in our toy model scenario. \textit{Top panel}: Empirical mean of the $\hat{y}_0$ residuals as a function of the CIB amplitude parameter $A_{\mathrm{CIB}}$. The standard MMF is significantly biased, with the bias increasing with $A_{\mathrm{CIB}}$, whereas the scMMF remains unbiased throughout. \textit{Middle panel}: Analogous plot for the signal-to-noise residuals. \textit{Bottom panel}: Signal-to-noise standard deviation. As expected, it is consistent with unity for both the MMF and the scMMF.}
\label{fig:toy}
\end{figure}

Figure \ref{fig:toy} shows the results of this idealised numerical experiment. The top panel shows the empirical fractional bias on $\hat{y}_0$, $\left\langle (\hat{y}_0 - y_0) \right\rangle /y_0 $, where we note that the input value of $y_0$ is $y_0 = 1.05 \times 10^{-4}$, and where angular brackets denote ensemble averaging over all our simulated observations for each value of $A_{\mathrm{CIB}}$. It is apparent that the standard MMF leads to a biased estimate of $y_0$ (blue data points), with the magnitude of the bias increasing with the amplitude of the foreground, as expected. On the other hand, also as expected, the scMMF leads to unbiased results (orange data points), as the foreground is perfectly deprojected. The error bars, in this and the other panels, are obtained by bootstrapping. The middle panel is an analogous plot for the bias in the signal-to-noise, $\left\langle q - \bar{q}_{\mathrm{t}} \right\rangle $, where  $\bar{q}_{\mathrm{t}}$ is the true signal-to-noise. As with $\hat{y}_0$, $q$ is biased with respect to the true signal-to-noise for the MMF, but it remains completely unbiased for the scMMF. Finally, the bottom panel shows the empirical standard deviation of the $q$ measurements, which for both MMFs is consistent with unity throughout, as expected.

As noted in Section \ref{subsec:formalism}, deprojection inevitably comes with a loss of signal-to-noise. In our toy model scenario, the true signal-to-noise values are $\bar{q}_{\mathrm{MMF}} = 13.05$ and $\bar{q}_{\mathrm{scMMF}} = 12.99$ for the MMF and the scMMF, respectively. This constitutes a signal-to-noise loss of about $0.5$\,\%. We recall that this penalty is independent of the amplitude of the foreground to be deprojected, i.e., in this case, from the value of $A_{\mathrm{CIB}}$. We remark, however, that it depends on the shape of the SED that is deprojected.

\subsection{Accounting for errors in the deprojection CIB: moment expansion}\label{subsec:moment}
%------------------------------
The SED of foregrounds such as the CIB is not perfectly well determined. Even if it were, variations of the SED along the line of sight and across the sky can translate into a mean SED that differs from the fundamental SED at any redshift and sky location. The moment expansion approach of \citet{Chluba2017} offers a way to account for these effects.
It is based on the idea that averaging processes (along the line of sight, within the instrument beam, or as part of the analysis) can be approximately captured using a Taylor series around suitable spectral pivot parameters describing the SED. This naturally leads to a list of spectral parameters and SED shapes that can be perturbatively added. 

In this work, we will consider the moment expansion to linear order. Let $I_{\nu} (\mathbf{p} (s,\bmath{\theta})) = A(s) \hat{I}_{\nu} (\mathbf{p} (s,\bmath{\theta}))$ be the SED of a CIB voxel at sky position $\bmath{\theta}$ and line-of-sight location parametrised by the affine parameter $s$. Here, $\hat{I}_{\nu} (\mathbf{p} (s,\bmath{\theta}))$ is the known as the fundamental SED, which is described by the spectral parameters $\mathbf{p}$, and $A(s)$ quantifies the contribution of each voxel to the total intensity. To linear order, the fundamental SED can be expanded around a pivot $\bar{\mathbf{p}}$ as \citep[see, e.g.,][]{Chluba2017,Rotti2021, Vacher2022a}
%-----------------------
\begin{equation}
\hat{I}_{\nu} (\mathbf{p} (s,\bmath{\theta})) \simeq \hat{I}_{\nu} (\bar{\mathbf{p}} (s,\bmath{\theta})) + ( \mathbf{p} - \bar{\mathbf{p}}) \cdot \nabla_{\bar{\mathbf{p}}} \hat{I}_{\nu} (\bar{\mathbf{p}}).
\end{equation}
%-----------------------
The line-of-sight and beam-averaged total intensity can then be written as
%-----------------------
\begin{equation}
\langle I_{\nu} (A, \mathbf{p}) \rangle \simeq  I_{\nu} (\bar{A},\bar{\mathbf{p}}) + \bmath{\omega} \cdot \nabla_{\bar{\mathbf{p}}} I_{\nu} (\bar{\mathbf{p}}),
\end{equation}
where $\bmath{\omega}$ contains the first-order moment coefficients of the expansion. That is, to linear order, the total integrated SED can be written as the fundamental SED evaluated at the pivot $\bar{\mathbf{p}}$ plus a moment-weighted sum of the first-order derivatives of the fundamental SED with respect to its spectral parameters. If CIB deprojection is desired, for a given choice of pivot $\bar{\mathbf{p}}$, one can deproject these first-order moment spectral shapes in addition to $I_{\nu} (\bar{A},\bar{\mathbf{p}})$. Doing this will mitigate the impact of deviations between the SED assumed in the deprojection and the true mean SED. One can also deproject higher-order moments, which account for foreground variation across the sky. However, since in this work we analyse small patches of the sky individually, we will consider only deprojection of the first-order moments.

The moment expansion method was first developed to describe the average SZ signal from realistic cluster atmospheres, including both temperature and velocity variations \citep{Chluba2013SZmoments}. It has also been successfully applied to the component separation problem of $B$-modes \citep{Mangilli2021, Remazeilles2021, Azzoni2021,Vacher2022a, Vacher2022}, the mapping of the relativistic SZ temperature \citep{Remazeilles2020}, and primordial CMB spectral distortions \citep{Rotti2021}.

% addition, the variation of the SED across the sky may be significant enough to have to be taken into account in a good description of the foreground. The moment expansion approach of \citet{Chluba2017} offers a way dealing of with these effects.

%Let us consider a foreground (e.g., the CIB) SED $\mathbfit{b} (\mathbfit{p})$, where $\mathbfit{p}$ is a number of parameters on which it depends. For the CIB, these can be, e.g., the emissivity index $\beta$ and the dust temperature at $z=0$, $T_0$. Let us assume that the mean SED is well described by $\mathbfit{b} (\bar{\mathbfit{{p}}})$, where $\bar{\mathbfit{{p}}}$ is some point in parameter space.

\section{MMF implementation: \texttt{SZ\lowercase{i}F\lowercase{i}}}\label{sec:implementation}

In this work we use the iMMF and sciMMF as implemented in \texttt{SZiFi}, the Sunyaev-Zeldovich iterative Finder\footnote{Pronounced `Sci-Fi'.}, a publicly-available Python package\footnote{\texttt{github.com/inigozubeldia/szifi}} which we have enhanced with a spectrally constrained mode in order to allow for foreground deprojection. A detailed description of \texttt{SZiFi} can be found in Section 4 of \citet{Zubeldia2022}. We use \texttt{SZiFi} with exactly the same configuration as in \citet{Zubeldia2022}, and so refer the reader to that work for the implementation details. However, for completeness, here we give a brief summary of how \texttt{SZiFi} operates.%, with a special focus on the details concerning foreground deprojection.

As input, \texttt{SZiFi} takes a set of intensity maps at different frequencies defined on the sphere (following the HEALPix scheme, \citealt{Gorski2005}), as well as a Galactic and a point source mask. As the final output, it produces a cluster catalogue containing values for the clusters' signal-to-noise, central Compton-$y$ $y_0$, angular scale $\theta_{500}$, and sky coordinates. \texttt{SZiFi} can be run in a non-iterative mode, in which the noise covariance is taken to be equal to the data covariance, and in an iterative mode, in which the detections from the first non-iterative step are masked out and the noise covariance is re-estimated. In addition, it can be run in a `cluster-finding' mode, in which clusters are detected blindly, and in a `fixed' mode, in which the values of the signal-to-noise and $y_0$ are extracted at the sky locations and angular scales of an input cluster catalogue. Finally, as a new feature, \texttt{SZiFi} can be run assuming either the standard MMF or the spectrally constrained MMF presented in Section \ref{sec:theory}, allowing for the deprojection of foregrounds with given SEDs.

\texttt{SZiFi} starts by tessellating the sky into a set of HEALPix pixels, which ensures that each point in the sky is covered only once. We use $N_{\mathrm{side}} = 8$, which leads to a total of 768 tiles. Then, square cut-outs of the frequency maps fully covering each tile are extracted as equirectangular projections of the original HEALPix frequency maps, centred at the centre of each tile. For each tile, the noise covariance is estimated as the cross-frequency power spectra of the frequency cut-outs using the MASTER algorithm \citep{Hivon2002} as implemented in \texttt{pymaster} \citep{Alonso2019}, properly dealing with the mode-coupling due to the Galactic mask, and having previously inpainted the regions zeroed by the point-source mask. An MMF is then constructed assuming a spatial template for the cluster tSZ signal. As in \citet{Zubeldia2022}, we assume the template due to the pressure profile of \citet{Arnaud2010}. This matched filter, which we recall can be either the standard MMF or our spectrally constrained MMF, can be evaluated either at a set of `search' angular scales $\theta_{500}$, in the cluster-finding mode, or at the angular scales and sky locations of an input cluster catalogue, in the fixed mode. In the fixed mode, \texttt{SZiFi} produces a $\hat{y}_0$ and a signal-to-noise map for each of the angular scales in the input catalogue and extracts the values at the corresponding cluster sky locations. In the cluster-finding mode, \texttt{SZiFi} produces a set of signal-to-noise maps, one for each search angular scale, and detections are then identified as the peaks of the three-dimensional signal-to-noise distribution (across sky location and angular scale) above a threshold of $q_{\mathrm{th}} = 5$. As in \citet{Zubeldia2022}, we consider 15 values of search angular scales logarithmically spaced between $\theta_{500}=0.5$\,arcmin and  $\theta_{500}=15$\,arcmin.

Finally, in the cluster-finding mode, detections within a given angular distance of each other are merged as a single one using a friends-of-friends algorithm, yielding a final, survey-wide non-iterative catalogue. Here, as in \citet{Zubeldia2022}, we use a merging distance of 10\,arcmin. In the fixed mode, on the other hand, detections are not merged. If iterative noise covariance estimation is desired, a mask in which the pixels within $3\theta_{500}$ of the detections in the non-iterative catalogue are zeroed is constructed and used in order to update the estimate of the noise covariance. This leads to an updated MMF and, in turn, to an iterative noise catalogue. Both the fixed and the cluster-finding modes support iterative covariance estimation. As shown in \citet{Zubeldia2022}, one single iteration suffices in order to eliminate the biases caused by the contamination of the noise covariance by the detections. In this work, we will always consider iterative MMFs and catalogues.

Finally, we note that, as a new feature, \texttt{SZiFi} can also be used to detect point sources in individual frequency maps. In order to do this, the matched filter spatial template is taken to be the instrument beam, instead of the tSZ signal due to a given pressure profile. As we detail in Section~\ref{sec:sims}, we use \texttt{SZiFi} as a point-source finder in order to construct a point-source mask, which is in turn used by \texttt{SZiFi} for cluster detection.

%As in \citet{Zubeldia2022}, our set of search angular scales is given by a set of 15 values of $\theta_{500}$ logarithmically spaced between $\theta_{500}=0.5$\,arcmin and  $\theta_{500}=15$\,arcmin. Although \texttt{SZiFi} allows for them, we do not use a Galatic or a point source mask, i.e., we consider the full sky.

%The Galactic and point source mask are then applied, and the pixels masked out by the point source mask are inpainted using a diffusive algorithm. 

\section{Simulated observations}\label{sec:sims}

In this section we describe the mock \textit{Planck}-like data we then analyse in Section \ref{sec:results}. We first describe our simulated maps and point-source mask, and then we detail the cluster catalogues that we obtain by applying \texttt{SZiFi} to them.

\subsection{Sky maps}

Except for the CIB, we use the same simulated maps that were employed in \citet{Zubeldia2022}. These consist of full-sky temperature maps observed in the six frequency channels of the \textit{Planck} High Frequency Instrument (HFI). The maps are pixelised following the HEALPix pixelisation scheme \citep{Gorski2005}, with $N_{\mathrm{side}} = 2048$, which is that used for the \textit{Planck} HFI legacy maps. Each of the six all-sky temperature maps contains the following components, all of which are convolved with the corresponding \textit{Planck} isotropic beam and the HEALPix pixel transfer function (except the instrumental noise, which is only convolved with the pixel transfer function):
%--------------------------------
\begin{itemize}
    \item \textbf{tSZ:} The Compton-$y$ map is taken from the \Websky simulation and is rescaled at each frequency by the non-relativistic tSZ SED. \Websky\footnote{\texttt{mocks.cita.utoronto.ca/websky}} \citep{Stein2019,Stein2020} is an all-sky second-order Lagrangian perturbation theory lightcone simulation with a minimum halo mass $M_{200} \sim 1.4 \times 10^{12} M_{\odot}$. Since the \Websky Compton-$y$ map has $N_{\mathrm{side}} = 4096$, we degrade it to $N_{\mathrm{side}} = 2048$ with the HEALPix \texttt{ud\_grade} function after beam convolution. We degrade in a similar way all the \Websky maps that we employ (kSZ, lensed CMB, and CIB). We neglect the effect of relativistic temperature corrections to the tSZ signal \citep[e.g.,][]{Itoh1998, Challinor1998, Sazonov1998, Chluba2012SZpack}, noting the this can in principle be added using the temperature--$y$ scaling relations from \citet{Lee2020,Lee2022}.
    \item \textbf{kSZ:} The kinetic SZ (kSZ) signal is also taken from the \Websky simulation, which includes both the late-time and the reionisation contributions \citep{Stein2020}.
    \item \textbf{Lensed CMB:} The lensed CMB is similarly taken from the \Websky simulation.
    \item \textbf{CIB:} The CIB is also taken from the \Websky simulation, which provides a map for each of the \textit{Planck} HFI frequencies. The \Websky CIB contains some bright point-like sources, which can contaminate cluster detection. Unlike in \citet{Zubeldia2022}, in which these point sources were removed by thresholding each of the CIB frequency maps, here we follow a more realistic approach in order to mask them, which we detail in Section \ref{masks}. At the \textit{Planck} frequencies, the \Websky CIB emission from redshift $z$ is given by a modified blackbody \citep{Stein2020}
    %---------------------
    \begin{equation}\label{eq:sed}
        I_{\nu} (z) = [\nu (1+z)]^{\beta} B_{\nu (1+z)} ( T_0 (1+z)^{\alpha}),
    \end{equation}
     %---------------------
   where $\nu$ is the observation frequency, $\beta = 1.6$ is the dust emissivity index, $T_0 = 20.7$\,K is the dust temperature at $z=0$, and $\alpha=0.2$, and where $B_\nu (T)$ is Planck's function. We note that we define the dust temperature parameter as $\beta_T = T_0^{-1}$.
    
    \item \textbf{Instrumental noise:} We generate a full-sky realisation of Gaussian white noise with noise levels, for increasing frequencies, of 77.4\,$\mu$K\,arcmin, 33\,$\mu$K\,arcmin, 46.8\,$\mu$K\,arcmin, 153.6\,$\mu$K\,arcmin, 46.8\,kJy\,sr$^{-1}$\,arcmin, and 43.2\,kJy\,sr$^{-1}$\,arcmin \citep{Planck2016viii}. The noise is uncorrelated between different channels, and, as noted above, is convolved only by the HEALPix pixel transfer function, and not by the instrument beam.  
\end{itemize}
%--------------------------------
As in \citet{Zubeldia2022}, our maps do not contain Galactic foregrounds and radio point sources. %\changeJ{However, these can to leading order be taken care of using appropriate sky-masks.}

\subsection{Point-source mask}\label{masks}

As we do not include Galactic foregrounds in our simulated data, we do not use a Galactic mask, i.e., we consider the full sky. We do, however, produce a point-source mask in order to tackle contamination from point-like sources present in the \Websky CIB. We construct such a mask by running \texttt{SZiFi} as a point-source finder on each of the six simulated frequency maps, producing a point-source catalogue for each frequency map. \texttt{SZiFi} is run in its iterative mode, and we impose a signal-to-noise selection threshold of 5. A final point-source catalogue is then produced by merging the individual frequency ones, and a binary point-source mask is constructed by zeroing all the pixels falling within a given masking distance of each detected point source. We use twice the beam FWHM of the channel in which each point source was detected as the masking distance. This point-source mask is used by \texttt{SZiFi} for cluster detection, (1) inpainting the masked regions with a diffusive algorithm for covariance estimation and for the construction of the MMF, and (2) not considering the masked regions of a buffered version of the mask in the peak-finding step. For more details about how this is done, we refer the reader to \citet{Zubeldia2022}.

\subsection{Cluster catalogues}\label{subsec:catalogues}

We apply \texttt{SZiFi} to our simulated maps, producing a number of cluster catalogues. As the CIB amplitude and spectral shape is very frequency-dependent, we explore the impact of the set of frequency channels used on the cluster observables. We also investigate the potential of deprojecting the first-order moments of the CIB SED to mitigate the impact of errors in the deprojection SED on the cluster observables. We note that, in addition to the (cluster-finding or fixed) noisy observables (signal-to-noise and $\hat{y}_0$), each catalogue also contains the true signal-to-noise of the detections, obtained by applying the corresponding MMF to the tSZ map only at the true cluster parameters. 

The catalogues that we obtain are the following:

%To this aim, we produce a number of catalogues, all of which are generated with the noise covariance estimated iteratively.

%---------------------
\begin{itemize}
    \item Three catalogues produced with the standard iMMF and \texttt{SZiFi} in the cluster-finding mode using, respectively, the four lowest \textit{Planck} HFI channels (100--353\,GHz), these channels in addition to the 545\,GHz channel, and all six HFI channels. Three additional catalogues are produced with the same configuration, but with \texttt{SZiFi} run in the fixed mode. For them, the impact of the CIB on the cluster observables can be observed in a cleaner way, free from other effects such as optimisation biases (see Section \ref{sec:results}). As the input catalogue in the fixed mode we use the \Websky halo catalogue with $M_{500} > 10^{14} M_{\odot}$, which contains most of the clusters that will by detected by our \textit{Planck}-like experiment (in addition to many others that will not be detected).
    \item Three additional iMMF cluster-finding catalogues with the CIB spatially randomised in the input maps. Specifically, for each of the tiles individually analysed by \texttt{SZiFi}, the CIB signal corresponding to another randomly-chosen tile is assigned, instead of the CIB associated to the tile. These catalogues allow us to assess whether a lack of tSZ--CIB spatial correlation  leads to any biases at all. We also produce three analogous catalogues with \texttt{SZiFi} in the fixed mode.
    \item A set of catalogues produced with our sciMMF and with \texttt{SZiFi} run in the cluster-finding mode for the three frequency combinations considered in the iMMF catalogues. We consider deprojection of the SED of the \Websky CIB and of two of its first-order moments, evaluating the SED and its derivatives at the true parameter values and at a fiducial redshift of $z=0.2$, which is representative of the \textit{Planck} cluster sample (see, e.g., \citealt{Planck2016xxvii}). In particular, for each frequency combination, we deproject:
    
    \begin{itemize}
        \item The \Websky CIB SED (`sciMMF mean catalogues').
        \item The CIB SED and its first-order moment with respect to the dust temperature parameter, $\beta_T = T_0^{-1}$ (`sciMMF mean+$\beta_T$ catalogues'). As noted in \citet{Chluba2017}, $\beta_T$ generally constitutes a better parametrisation than the temperature $T_0$.
        \item The CIB SED and its first-order moment with respect to the CIB emissivity index $\beta$ (`sciMMF mean+$\beta$ catalogues').
        \item The CIB SED and its first-order moments with respect to both $\beta_T$ and $\beta$ (`sciMMF mean+$\beta_T$+$\beta$ catalogues').
    \end{itemize}
%---------------------
An analogous set of catalogues is obtained with \texttt{SZiFi} in the fixed mode, leading to a total of 24 sciMMF catalogues.

\end{itemize}

Finally, we note that in Section \ref{subsec:robustness} we investigate the robustness of our results to the choice of deprojection SED. In order to do so, we produce an number of additional cluster catalogues, which are described in that section.

\section{Results and discussion}\label{sec:results}

In this section we analyse our mock \textit{Planck}-like catalogues, studying the impact of the cluster-correlated CIB on the cluster observables and on the survey completeness function, and assessing the capabilities of CIB deprojection with our sciMMF to mitigate these CIB-induced effects.

\subsection{iMMF: Bias on signal-to-noise, completeness, and number counts}\label{subsec:immf}

\subsubsection{CIB-induced bias on the signal-to-noise}\label{subsec:bias_snr}

\begin{figure*}
\centering
\includegraphics[width=0.7\textwidth,trim={00mm 6mm 0mm 15mm},clip]{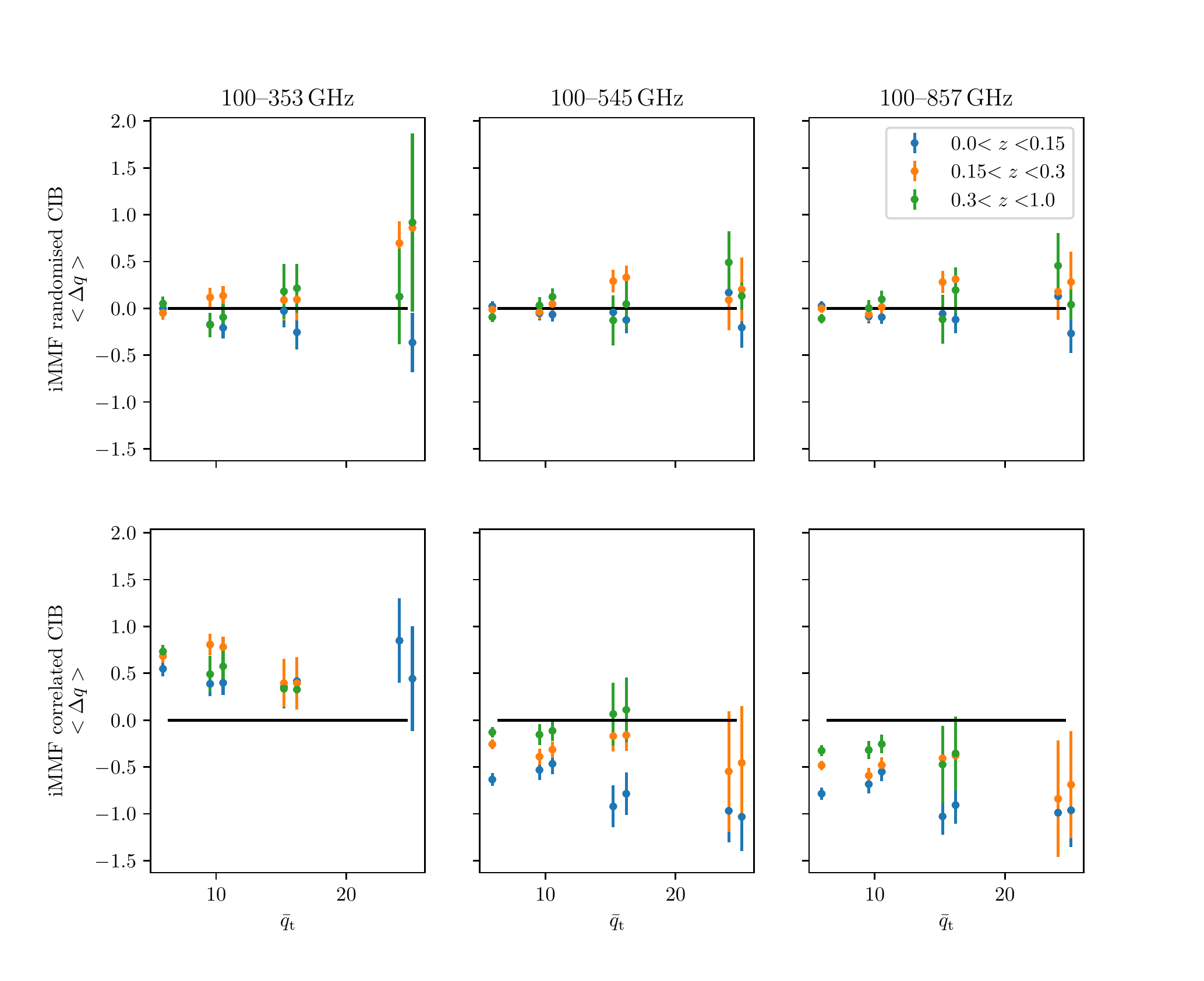}
\caption{Mean of the signal-to-noise residuals for our iMMF catalogues, both if the CIB is spatially correlated with the tSZ signal (as it would in a real experiment; lower panels) and if it is randomised (upper panels). A significant frequency and redshift-dependent bias is seen in the former case, whereas no bias is seen in the latter. The data points on the left-hand side of each signal-to-noise bin correspond to the catalogues obtained with \texttt{SZiFi} applied in its fixed mode, with the signal-to-noise extracted at the clusters' true parameters (angular and sky location), whereas those on the right-hand side of each bin correspond to the blind cluster catalogues obtained with \texttt{SZiFi} in its cluster-finding mode.}
\label{fig:corrvsuncorr_q}
\end{figure*}

Figure \ref{fig:corrvsuncorr_q} shows the empirical mean of the signal-to-noise residuals $\Delta q$ for our iMMF and randomised iMMF cluster catalogues (lower and upper panels, respectively) binned as a function of true signal-to-noise and redshift. We use four signal-to-noise bins logarithmically spaced between $\bar{q}_{\mathrm{t}} = 5$ and $\bar{q}_{\mathrm{t}} = 30$ and three redshift bins with edges given by $z=0$, $z = 0.15$, $z=0.3$, and $z=1$. For each bin, the empirical mean of $\Delta q$ is shown for both the cluster-finding and fixed cases (data points on the left-hand and right-hand side of the centre of each bin, respectively); the error bars are obtained by bootstrapping. In the fixed case, in which the signal-to-noise $q$ is extracted at the true cluster parameter values ($\theta_{500}$ and sky location), the residuals are defined as $\Delta q = q - \bar{q}_{\mathrm{t}}$, where $\bar{q}_{\mathrm{t}}$ is the cluster true signal-to-noise. On the other hand, in the cluster-finding case, in which the clusters are detected blindly, the signal-to-noise residuals $\Delta q$ are defined as $\Delta q = q_{\mathrm{opt}} -  (\bar{q}_{\mathrm{t}}^2 + 3)^{1/2}$, where $q_{\mathrm{opt}}$ is the detection or `optimal' signal-to-noise. The definition of $\Delta q$ for blind detections accounts for the optimisation bias caused by signal-to-noise maximisation over three parameters in a noisy data set, as discussed extensively in \citet{Zubeldia2021}.

It is apparent in Figure \ref{fig:corrvsuncorr_q} that if the CIB is not spatially correlated with the tSZ signal (upper panels), the empirical mean of the signal-to-noise residuals is consistent with zero across the signal-to-noise and redshift bins: the signal-to-noise is unbiased. This was already seen in \citet{Zubeldia2022}, where the CIB was also randomised, and was to be expected from a theoretical point of view. Indeed, we are applying a linear operation (matched filtering) to a random field (the CIB) and extracting the value of the filtered field (the signal-to-noise fluctuation field due to it, $\Delta q_{\mathrm{CIB}}$) at locations that are uncorrelated with the field itself. This procedure provides an unbiased set of samples of the filtered field, i.e., an unbiased set of $\Delta q_{\mathrm{CIB}}$ samples. Therefore, the mean of these samples will be consistent with the true mean of the filtered field. This is zero, as \texttt{SZiFi} discards the $\mathbfit{l} = 0$ mode. (We note that, in reality, matched filtering is not a linear operation on the data if the covariance is estimated from the data itself, but this is a second-order effect in the iterative approach, see \citealt{Zubeldia2022}.) %really?

On the other hand, the lower panels of Figure \ref{fig:corrvsuncorr_q} show that if there is spatial correlation between the CIB and the tSZ signals, a significant bias in the signal-to-noise is observed. This bias is at the level of about $0.5\,\sigma$ \emph{per cluster}, as the signal-to-noise has (approximately) unit standard deviation (see Section \ref{subsec:std}). The existence of this bias was also to be expected, as now the values of the signal-to-noise fluctuation field $\Delta q_{\mathrm{CIB}}$ are extracted at locations that are correlated with the field itself, i.e., they do not constitute an unbiased sample. This bias, which hereafter we refer to as the `CIB bias', can be seen to depend strongly on the frequency channels used in the analysis, as well as on redshift, but less so on the cluster's true signal-to-noise at fixed redshift. %interpretation

We note that in the cluster-finding case we do not show the residuals for the lowest signal-to-noise bin, as they are biased (high) due to Malmquist bias. Indeed, the cluster-finding catalogues are constructed by imposing a signal-to-noise threshold at $q_{\mathrm{opt}} = 5$: near the threshold, at fixed $\bar{q}_{\mathrm{t}}$, clusters with a preferentially high value of $q_{\mathrm{opt}}$ are detected. Malmquist bias, however, does not affect the fixed catalogues, which contain the signal-to-noise of all the clusters in the input catalogue, regardless of whether they fall below the selection threshold or not. We note that Malmquist bias is only relevant below $\bar{q}_{\mathrm{t}} \simeq 7$, as the signal-to-noise has approximately unit standard deviation. In addition, Malmquist bias would be of no concern in a likelihood analysis, in which it would be naturally accounted for if the sample selection is properly modelled via the completeness function. %explain that we previously select a c

%Compare with J-B bias

\subsubsection{Completeness}\label{subsec:completeness}

The CIB bias seen in Figure \ref{fig:corrvsuncorr_q} has an impact on the survey completeness function. The completeness of a survey is defined as the probability for a cluster to be included in the catalogue at given true cluster parameters (e.g., mass and redshift, or true signal-to-noise). Assuming the signal-to-noise residuals to follow a Gaussian distribution with zero mean and unit standard deviation (see Section \ref{subsec:std} for the latter), the completeness at given true signal-to-noise, $P(\mathrm{i}|\bar{q}_{\mathrm{t}})$, where i denotes the inclusion of the cluster in the catalogue, can thus be written as
%-------------------------
\begin{equation}\label{erf}
P(\mathrm{i}|\bar{q}_{\mathrm{t}}) = \frac{1}{2} \left[ 1 - \mathrm{erf} \left( \frac{q_{\mathrm{th}} - \bar{q}}{\sqrt{2}} \right) \right].
\end{equation}
%-------------------------
Here, $q_{\mathrm{th}}$ is the signal-to-noise selection threshold, and $\bar{q} = \bar{q}_{\mathrm{t}}$ for the fixed case and $\bar{q} = (\bar{q}_{\mathrm{t}}^2 + 3)^{1/2}$ for the cluster-finding case. We recall that the selection threshold that we use is $q_{\mathrm{th}} = 5$ for the cluster-finding catalogues. As noted in Section \ref{subsec:bias_snr}, in the fixed mode, signal-to-noise measurements are obtained for all the clusters in the input catalogue. In this section, however, in order to investigate the completeness in the fixed case too, we also impose a threshold $q_{\mathrm{th}} = 5$ on the fixed catalogues, selecting the clusters whose signal-to-noise falls above it.

\begin{figure*}
\centering
\includegraphics[width=0.7\textwidth,trim={00mm 5mm 0mm 10mm},clip]{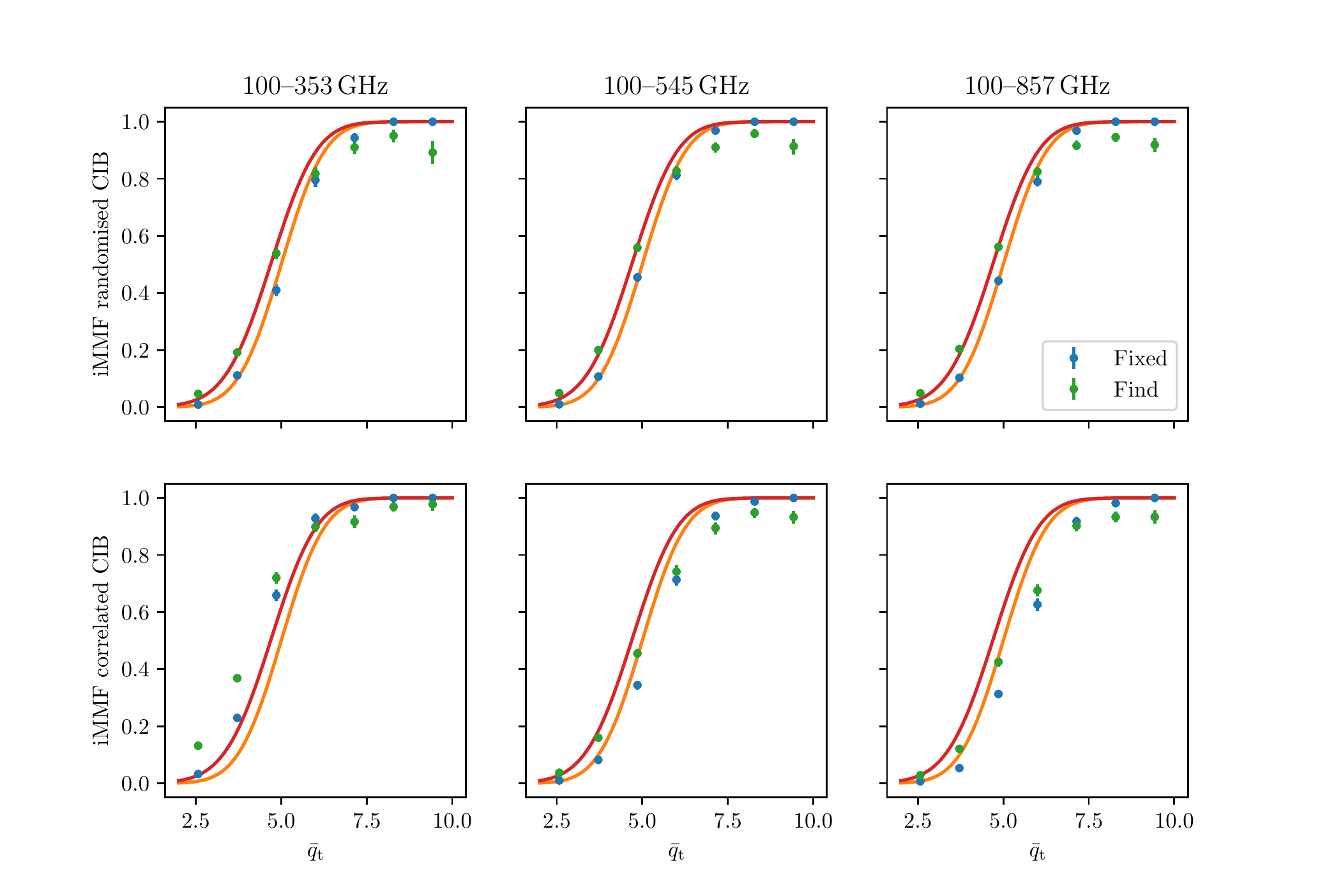}
\caption{Survey completeness for our iMMF catalogues, both if the CIB is correlated with the tSZ signal (lower panels) and if it is randomised (upper panels). The blue and green data points correspond to the empirical completeness of our fixed and cluster-finding catalogues, respectively, and the orange and red curves to the corresponding theoretical predictions, given by Eq. (\ref{erf}). For the fixed catalogues, the empirical completeness agrees with its theoretical expectation if the CIB is randomised, but they differ if the CIB--tSZ correlation is included, as a consequence of the CIB bias. For the cluster-finding catalogues, an additional bias due to cluster confusion is observed in all the panels (see Appendix \ref{appendix:confusion}).}
\label{fig:completeness}
\end{figure*}

Figure \ref{fig:completeness} shows the empirically-estimated completeness as a function of $\bar{q}_{\mathrm{t}}$ for our iMMF and randomised iMMF catalogues (i.e., the same catalogues that are shown in Figure \ref{fig:corrvsuncorr_q}). The data points corresponding to the fixed and cluster-finding catalogues are shown in blue and green, respectively. Their associated theoretical predictions, computed with Eq. (\ref{erf}), are shown in orange and red, respectively. The error bars are calculated by bootstrapping; for each bin, they correspond to the 15.9\,\% and 84.1\,\% bootstrapped quantiles, taking into account the asymmetric nature of the data (the completeness cannot take negative values or values greater than unity). The fixed catalogues offer a cleaner view of the impact of the CIB bias on the completeness. If the CIB is randomised (upper panels, blue data points), the empirical completeness is fully in agreement with its theoretical prediction (orange curves). This is not the case if the CIB is correlated with the tSZ signal (lower panels, blue data points): the empirical completeness is biased high (low) with respect to its theoretical expectation for $100$--$353$\,GHz ($100$--$545$\,GHz and $100$--$857$\,GHz). This bias is an obvious consequence of the CIB bias in the signal-to-noise shown in Figure~\ref{fig:corrvsuncorr_q}: at given true signal-to-noise, if the observed signal-to-noise is biased high (low), more (fewer) clusters than expected will be detected, and therefore the real completeness function will be above (below) its theoretical expectation.

The cluster-finding case is more complex, as in addition to the CIB-induced bias, the completeness suffers from an additional bias caused by cluster confusion: clusters which are too close to each other to be distinguished due to the finite size of the \textit{Planck} beams are detected as single objects. This $\sim 5 \%$ effect is discussed in more detail in Appendix \ref{appendix:confusion}.

\subsubsection{Number counts}\label{subsec:number_counts}

\begin{figure*}
\centering
\includegraphics[width=0.7\textwidth]{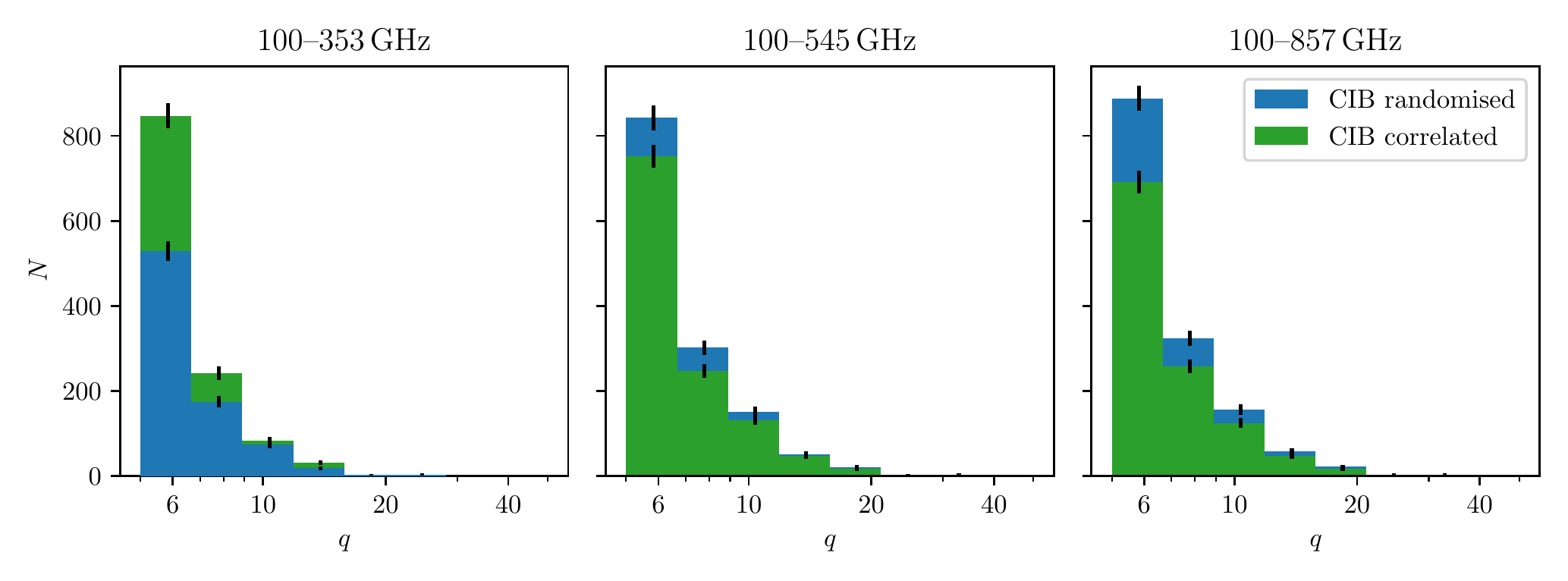}
\caption{Cluster number counts as a function of signal-to-noise for our iMMF catalogues, both if the CIB--tSZ correlation is included (blue bars), and if it is removed by randomising the CIB tiles (green bars). For the three frequency combinations considered, the difference in the number counts between the correlated and randomised catalogues is larger than the associated Poisson error bars, which are shown in black, for several of the bins.}
\label{fig:histograms}
\end{figure*}

As well as biasing the completeness, the CIB bias has a non-negligible impact on the number counts. Figure \ref{fig:histograms} shows the cluster number counts as a function of detection, or optimal, signal-to-noise $q_{\mathrm{opt}}$ for our iMMF and random iMMF cluster-finding catalogues (green and blue bars, respectively). In order for the number counts to be closer to the real \textit{Planck} ones, we have applied the \textit{Planck} $60$\,\% Galactic mask (available in the PLA) to our catalogues, which we recall are obtained for the full sky. For each frequency combination, the difference between the randomised and the correlated number counts is simply a consequence of the CIB bias seen in Figure \ref{fig:corrvsuncorr_q}. Indeed, more (fewer) clusters are detected in the correlated case relative to the randomised case for $100$--$353$\,GHz ($100$--$545$\,GHz and $100$--$857$\,GHz), as the CIB bias takes positive (negative) values. We note that for all three frequency combinations, the difference in the number counts is larger than the associated Poisson errors (shown in black) in several of the bins. In a likelihood analysis, it is the randomised number counts that would be predicted if the halo mass function and the tSZ signal are modelled accurately, but the CIB--tSZ correlation is neglected. On the other hand, the correlated number counts are those that would be obtained in a real observation. Given that the difference between the two is statistically significant, the CIB bias could potentially bias cosmological inference. This justifies the application of our sciMMF to our \textit{Planck}-like data, which we discuss in in Section \ref{subsec:sciMMF}.

The CIB-induced number-count change that we observe is larger than what is reported in \citet{Melin2018}. In that study, a $\sim\,10$\,\% decrement in the number of detections between $z=0.4$ and $z=0.8$ was observed by injecting clusters with dust emission to real \textit{Planck} data and applying an MMF using all the \textit{Planck} HFI channels. The reason for this discrepancy probably lies in the dust emission models assumed in each simulated data set. We have used the \Websky simulation, for which the CIB power spectra and the CIB--tSZ cross-correlation are in agreement with \textit{Planck} observations, as is also the CIB--CMB lensing cross-correlation  \citep{Stein2019}. \citet{Melin2018}, on the other hand, generated their mock observations assuming a dust emission model that was fit to \textit{Planck} data. This model was, however, seen to underpredict the fixed aperture photometry flux at the locations of \textit{Planck}'s clusters for \textit{Planck}'s three highest-frequency channels (see Figure A.1 of \citealt{Melin2018}). We believe that this lower-amplitude injected dust emission is responsible for \citet{Melin2018} finding a smaller CIB-induced change in the retrieved number counts.

\subsubsection{CIB-induced bias on $y_0$}\label{subsec:y0}

The signal-to-noise $q$ is our preferred cluster observable, as it is the observable through which the sample is selected and it has well-behaved properties, particularly in the cluster-finding case, in which the optimisation bias can be modelled accurately with the simple prescription of \citet{Zubeldia2021}. However, it is also possible to consider the cluster amplitude parameter, $y_0$, which we recall is defined as the value of the cluster's central Compton-$y$ parameter. We study the impact of the CIB on this parameter in Appendix \ref{appendix:y0}, finding a $\sim 5$--$10\,\%$ CIB-induced bias.

\subsection{sciMMF: suppressing the CIB bias}\label{subsec:sciMMF}

\subsubsection{Bias on the signal-to-noise}\label{subsec:cmmfbias}

As discussed in Section \ref{subsec:immf}, the CIB--tSZ correlation introduces a bias in the matched filter signal-to-noise, which in turn translates into biases in the survey completeness function and the cluster number counts. Here, we investigate the capabilities of our sciMMF to tackle these effects.

\begin{figure*}
\centering
\includegraphics[width=0.8\textwidth,trim={00mm 15mm 0mm 20mm},clip]{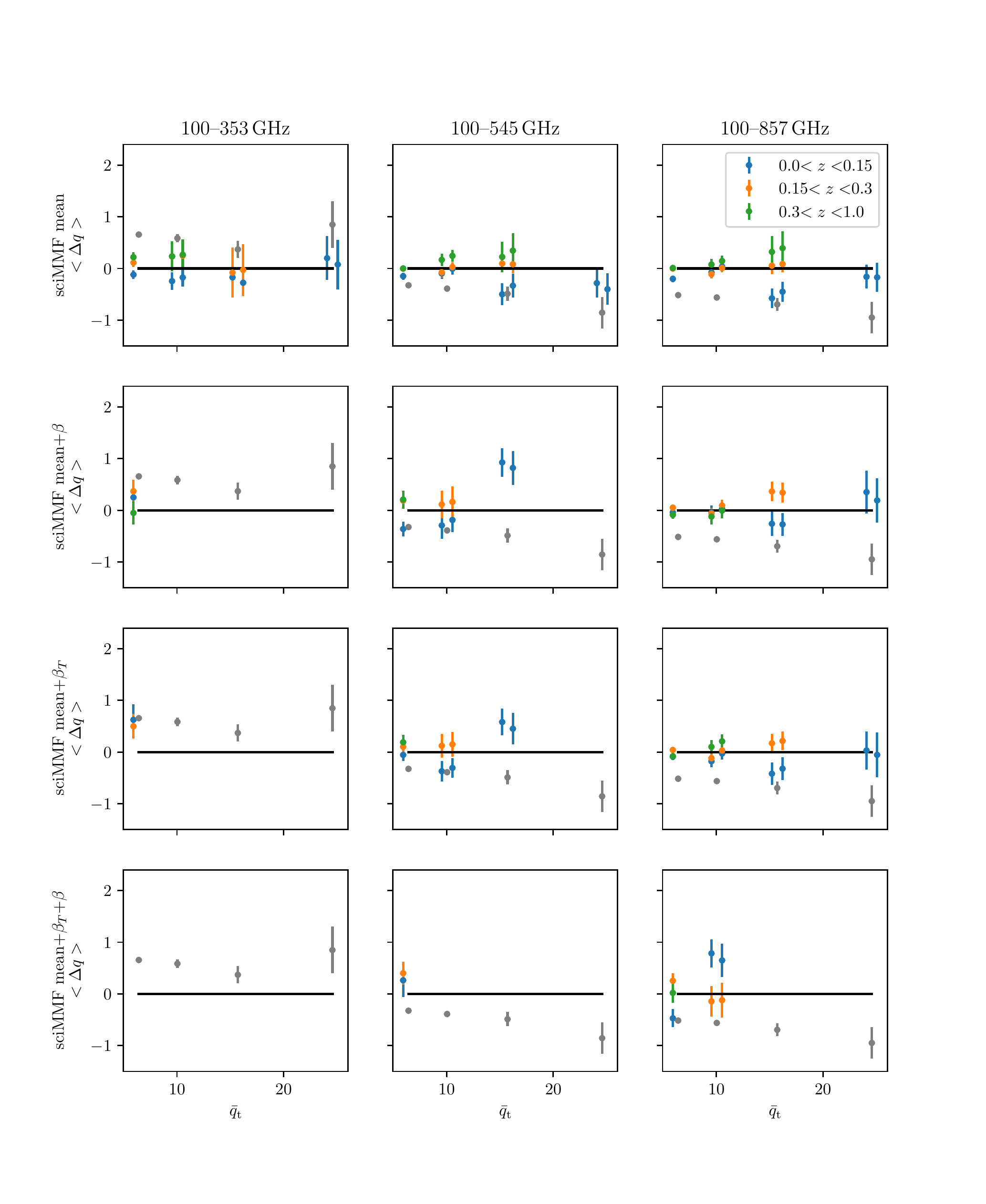}
\caption{Empirical mean of the signal-to-noise residuals binned in true signal-to-noise $\bar{q}_{\mathrm{t}}$ and redshift for our simulated sciMMF cluster catalogues. The deprojection SEDs are: the CIB SED (first row), the CIB SED and its first-order moment with respect to the dust emissivity index $\beta$ (second row), the CIB SED and its first-order moment with respect to the dust temperature parameter $\beta_T=T_0^{-1}$ (third row), and the CIB SED and its first-order moments with respect to both $\beta$ and $\beta_T$ (fourth row). For comparison, the mean of the signal-to-noise residuals for our iMMF cluster catalogues, averaged in a single redshift bin, are also shown (grey data points). We note that only the data points for bins containing more than five clusters are shown.}
\label{fig:dep_q}
\end{figure*}

Figure \ref{fig:dep_q} shows the empirical mean of the signal-to-noise residuals $\Delta q$ of our sciMMF cluster catalogues binned in true signal-to-noise and redshift, for both the fixed and cluster-finding cases (data points on the left-hand and right-hand side of the centre of each bin, respectively). In addition, for comparison with the standard iMMF, in grey we show the empirical mean of the residuals for our three iMMF catalogues, averaged over all redshifts. This is the same data that is shown shown on the left-hand side of the bins in the lower panels of Figure \ref{fig:corrvsuncorr_q}, but binned in a single redshift bin. As noted in Section \ref{subsec:catalogues}, the four SED combinations that we consider for deprojection are: the CIB SED (first row of Figure \ref{fig:dep_q}), the CIB SED and its first-order moment with respect to the emissivity index $\beta$ (second row), the CIB SED and its first-order moment with respect to the dust temperature parameter $\beta_T = T_0^{-1}$ (third row), and the CIB SED and its first-order moments with respect to both $\beta$ and $\beta_T$ (fourth row). We recall that the CIB SED and its derivatives are evaluated at the input parameter values of the \Websky simulation and at a fiducial redshift of $z=0.2$ (see Section \ref{subsec:catalogues}). The choice of fiducial redshift is representative of the \textit{Planck} cluster sample: although the CIB peaks at a higher redshift, it is the emission that is correlated with our detected clusters that we want to remove. We note that, as in Figure \ref{fig:corrvsuncorr_q}, we do not show the data point for the lowest signal-to-noise bin in the cluster-finding case, as it suffers from Malmquist bias. In addition, we only consider bins containing more than 5 clusters, so that the error bars are reasonably well-determined.

As it can be observed in Figure \ref{fig:dep_q}, deprojection significantly suppresses the CIB bias, which, we recall, for the standard iMMF is at the level of about $0.5$ in signal-to-noise, i.e., of about $0.5\,\sigma$ per cluster (grey data points). In particular, if only the mean CIB SED is deprojected, the signal-to-noise is completely unbiased for $z > 0.15$ for all three frequency combinations to the level of accuracy provided by our mock observations (except perhaps a small $\sim 0.1\,\sigma$ bias at $z > 0.3$ for $100$--$353$\,GHz). A small $\sim 0.1\,\sigma$ bias is observed at $z < 0.15$ for $100$--$545$\,GHz and $100$--$857$\,GHz, which we interpret as being caused by the shape of the CIB SED at $z < 0.15$ not being sufficiently close to the shape at $z = 0.2$, the fiducial redshift at which the deprojection SED is evaluated. 

If, in addition to the mean CIB SED, its first-order moment with respect to either $\beta$ or $\beta_T$ is deprojected, the signal-to-noise is completely unbiased for $100$--$854$\,GHz at all redshifts. This demonstrates the effectiveness of the moment expansion approach at mitigating the impact of errors in the assumed deprojection SED. For $100$--$545$\,GHz, deprojecting either of the first-order moments also leads to an unbiased signal-to-noise at $z > 0.15$, as does jointly deprojecting both first-order moments for $100$--$854$\,GHz. However, a bias is observed for $z < 0.15$, which we also attribute to the low-redshift CIB SED shape. Given that these three sciMMFs carry a very significant loss signal-to-noise relative to the iMMF (see Section \ref{subsec:snrpenalty}, and notice the size of the error bars), this is of no concern, as their use in practice would not be advised. The same can be said about the four remaining sciMMFs, whose associated signal-to-noise penalty is extreme (see Section \ref{subsec:snrpenalty}).

In summary, three of the sciMMFs retaining a significant signal-to-noise are virtually unbiased across signal-to-noise and redshift: the $100$--$854$\,GHz sciMMF deprojecting the CIB SED and its first-order moment with respect to $\beta$, the analogous one for $\beta_T$, and the $100$--$353$\,GHz sciMMF deprojecting only the mean CIB SED. The empirical mean of $\Delta q$ across the whole cluster sample for each of these sciMMFs is shown in Figure \ref{fig:robust} (central data points of both panels), where it can be seen that it is consistent with zero for the three sciMMFs. We thus take these three sciMMFs as our preferred ones, and investigate their robustness against the choice of wrong deprojection SED parameter values in Section \ref{subsec:robustness}.

As noted in Section \ref{subsec:y0}, although the signal-to-noise is our preferred cluster observable, one can also consider the amplitude parameter $y_0$. As we show in Appendix \ref{appendix:y0}, our sciMMFs are very effective at removing the CIB bias from $y_0$ as well, with the same sciMMFs that produce unbiased signal-to-noise measurements also delivering unbiased $y_0$ measurements.
 
\subsubsection{Standard deviation of the signal-to-noise}\label{subsec:std}

If the noise covariance is estimated correctly, the signal-to-noise has, by construction, unit standard deviation around its expected value, the mean signal-to-noise $\bar{q}_{\mathrm{t}}$. Note that by noise we mean all the signals in the data other than the tSZ signal of the detected clusters, including the CIB. For all our (sc)iMMFs, \texttt{SZiFi} estimates the noise covariance as the cross-frequency power spectra of the temperature cut-outs, which we recall are masked at the locations of the detections following \citet{Zubeldia2022}. This procedure assumes that the noise power spectrum is constant across each cut-out. If, however, the signal-to-noise is extracted at locations where the noise power is systematically higher than on average, then it will be more noisy than expected, i.e., its standard deviation will be greater than unity.

\begin{figure*}
\centering
\includegraphics[width=0.8\textwidth,trim={00mm 5mm 0mm 10mm},clip]{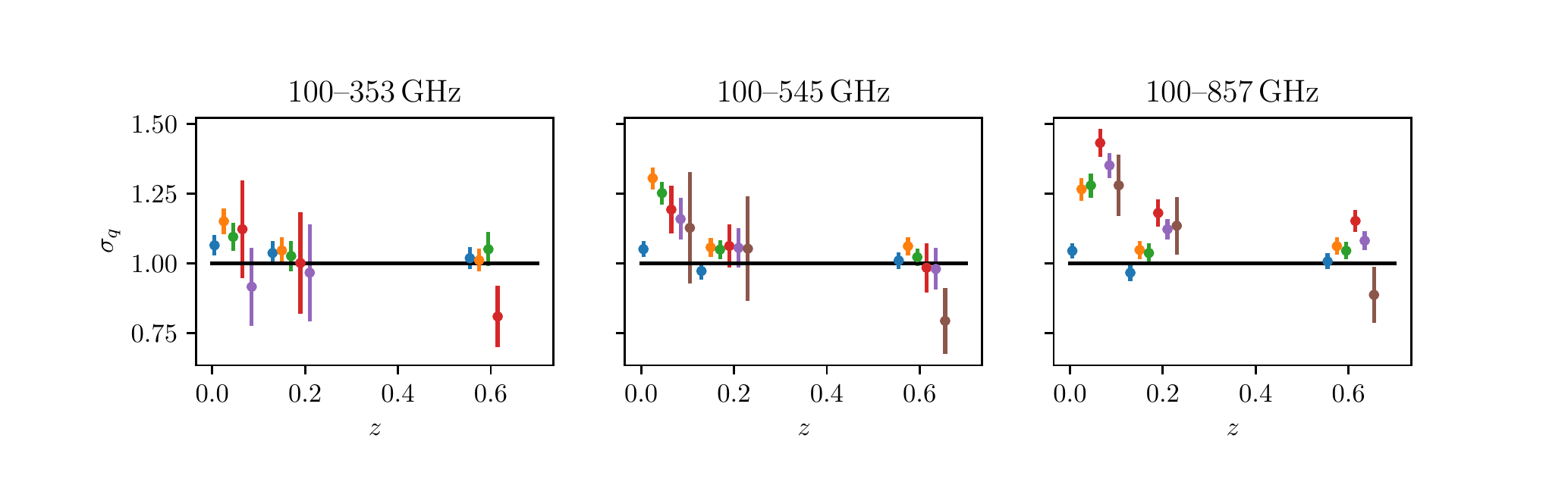}
\caption{Signal-to-noise standard deviation for all our fixed cluster catalogues, computed across three redshift bins with edges $z=0$, $z=0.15$, $z=0.3$, and $z=1$. For each redshift bin, the data points correspond, from left to right, to the catalogues obtained with the randomised iMMF (in blue), iMMF (in orange), sciMMF with deprojection of the mean CIB SED (in green), and sciMMF with additional deprojection of the first-order moments with respect to $\beta$, $\beta_T$, and both $\beta$ and $\beta_T$ (in red, purple, and brown, respectively).}
\label{fig:std}
\end{figure*}

Figure \ref{fig:std} shows the standard deviation of the signal-to-noise residuals $\Delta q$, $\sigma_q$, empirically estimated across three redshift bins with clusters with $\bar{q}_{\mathrm{t}} > 5$ for all our fixed catalogues. Similar results are obtained for our cluster-finding catalogues, which are not shown for the sake of conciseness. For each bin, the data points correspond, from left to right, to our random iMMF, iMMF, sciMMF with mean SED deprojection, and sciMMFs deprojecting the mean SED and the first-order moment with respect to $\beta$, $\beta_T$, and both $\beta$ and $\beta_T$. As expected, the standard deviation for the random iMMF catalogues is consistent with unity throughout. This is, however, not the case if CIB is correlated, especially in the lowest redshift bin, for which $\sigma_q \sim 1.2$--$1.4$ for the iMMF for all three frequency combinations. This indicates that the CIB has more power around the locations of clusters than on average, especially at low redshift. As well as depending on redshift, $\sigma_q$ is also seen to depend on the frequency combination of choice, increasing with the maximum frequency taken into account. This was to be expected, as within the \textit{Planck} frequency range, the CIB SED increases strongly with frequency, and as a consequence the CIB contributes more strongly to the total noise covariance at higher frequencies.

Deprojection generally brings $\sigma_q$ closer to unity relative to the standard iMMF case, but in some instances full agreement with unity is never achieved. This is a consequence of the deprojection SED not matching perfectly the real SED of the cluster-correlated CIB, with some correlated CIB power still remaining. Indeed, if the cluster-correlated CIB SED did not change with redshift and perfectly matched the deprojection SED, the CIB would be completely removed and would have no impact on $\sigma_q$ at all (see our toy model of Section \ref{subsec:toymodel}, in particular the lower panel of Figure \ref{fig:toy}).

In summary, the CIB--tSZ correlation systematically increases the signal-to-noise standard deviation over its expected value for our iMMF, as well as for our sciMMFs in some instances. In a number count analysis, this additional scatter in the observable could be modelled as part of the intrinsic scatter, which is typically introduced in order to accounts for the dispersion of the true tSZ signal around its mean value at fixed mass and redshift due to deviations from sphericity, etc. A quantification of how much this CIB-induced scatter affects the intrinsic scatter, as well as whether it can lead to any biases in the cosmological inference if unaccounted for, is beyond the scope of this paper.

\subsubsection{Completeness}

\begin{figure*}
\centering
\includegraphics[width=0.7\textwidth,trim={00mm 15mm 0mm 20mm},clip]{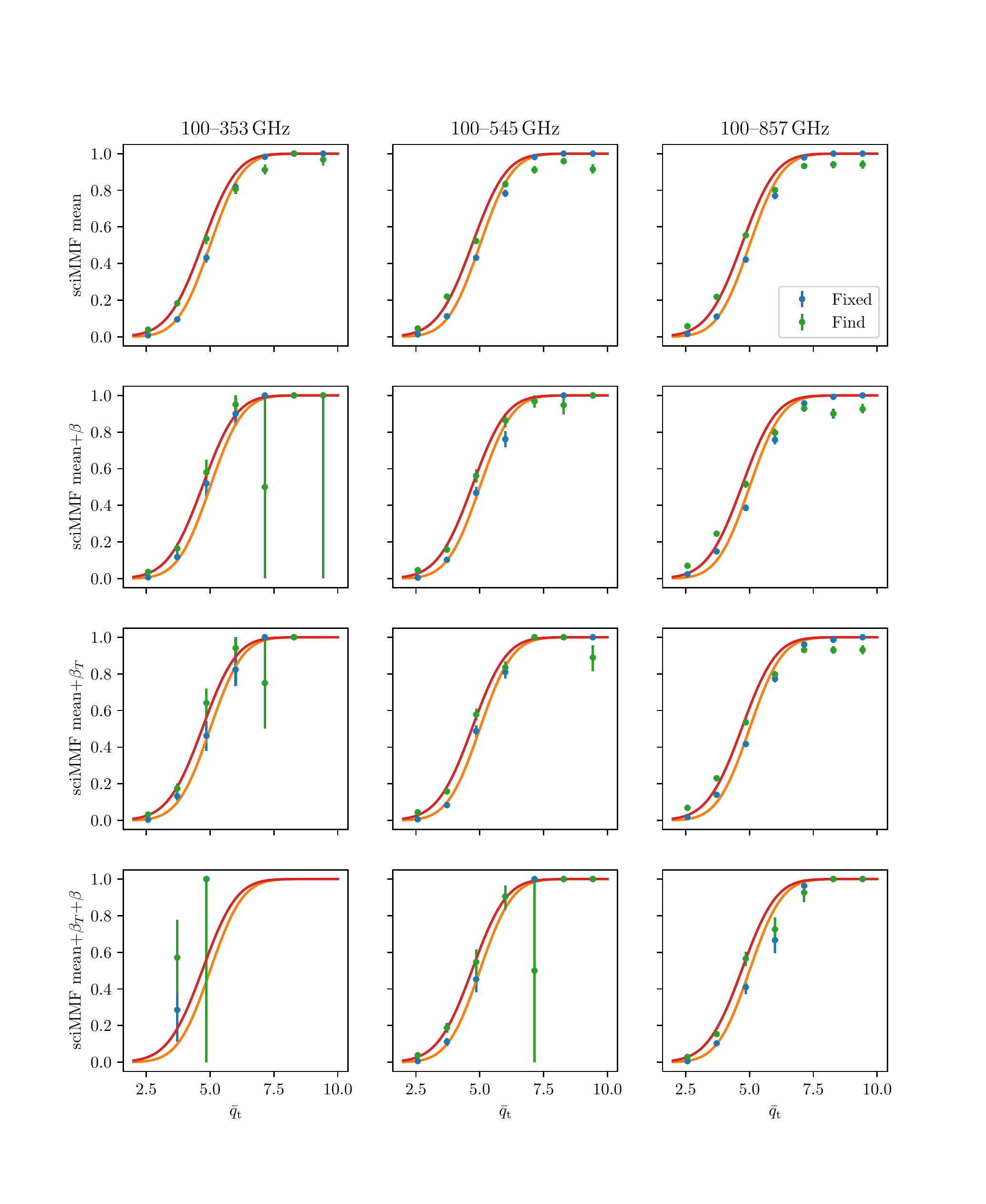}
\caption{Empirical survey completeness function for our sciMMF catalogues, both in the fixed and cluster-finding cases (blue and green data points, respectively), shown along their corresponding theoretical predictions (orange and red curves, respectively).}
\label{fig:dep_com}
\end{figure*}

The suppression of the CIB bias achieved by our sciMMFs has a positive impact on the survey completeness function, bringing it closer to its theoretical expectation. This effect can be seen in Figure \ref{fig:dep_com}, which shows the empirically-estimated completeness for our sciMMF catalogues both in the fixed and cluster-finding cases (blue and green data points, respectively), along with their respective theoretical predictions, given by Eq. (\ref{erf}) (orange and red curves, respectively). The error bars are obtained by bootstrapping and correspond to the 15.9\,\% and 84.1\,\% bootstrapped quantiles. For the fixed catalogues, the completeness is indeed much closer to its theoretical expectation than for the standard iMMF (see Figure \ref{fig:completeness}). However, in some instances for which no CIB bias is observed (e.g., for $100$--$857$\,GHz for deprojection of the mean CIB SED and the first-order moment with respect to either $\beta$ or $\beta_T$), the data points are slightly above the theoretical completeness for $\bar{q}_{\mathrm{t}} < 5$, and below it for $\bar{q}_{\mathrm{t}} > 5$ (second and third panels in the last column). This is a consequence of the signal-to-noise standard deviation $\sigma_q$ being slightly greater than unity (see Section \ref{subsec:std}). As a consequence, the real completeness is given by Eq. (\ref{erf}) but substituting $\sqrt{2}$ by $\sqrt{2} \sigma_q$, i.e., it is wider than expected. As discussed in Section \ref{subsec:std}, this effect could be accounted for through an intrinsic scatter term. 

In addition, for the sciMMFs with virtually no CIB bias, the cluster-finding completeness is seen to be about $\sim 5$\,\% low relative to its theoretically-predicted value for $\bar{q}_{\mathrm{t}} > 5$. This was also observed for the standard iMMF in Section \ref{subsec:completeness}, and is similarly caused by cluster confusion (see Appendix \ref{appendix:confusion}). At $\bar{q}_{\mathrm{t}} < 5$, this effect is observed to be compensated by the increase in the signal-to-noise standard deviation, with the empirical cluster-finding completeness generally lying slightly above the theoretical curve.

\subsection{Signal-to-noise penalty}\label{subsec:snrpenalty}

\begin{figure*}
\centering
\includegraphics[width=0.8\textwidth,trim={00mm 15mm 0mm 20mm},clip]{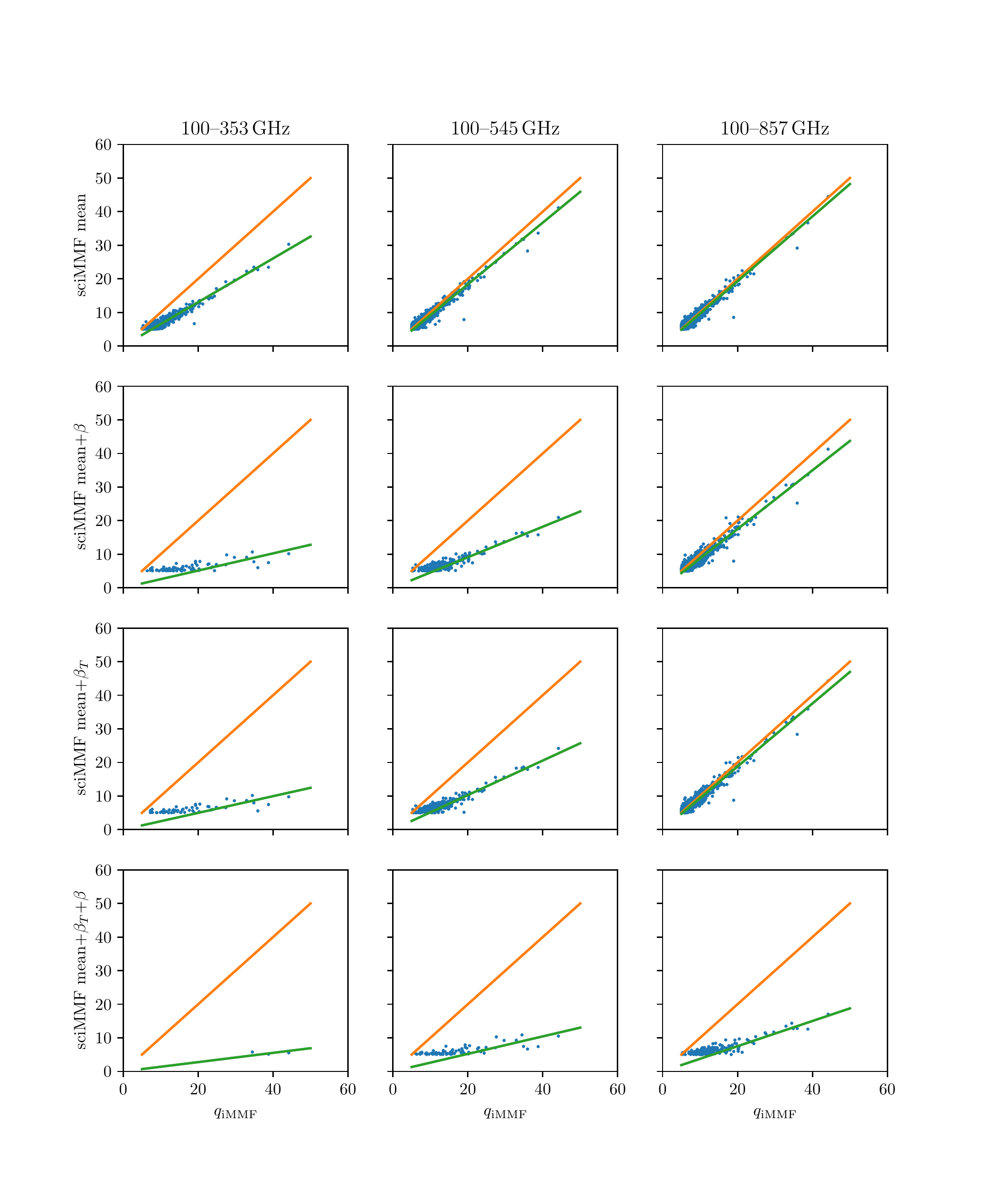}
\caption{Signal-to-noise measurements in our different sciMMF catalogues plotted against those in our $100$--$857$\,GHz random iMMF catalogue, which is the MMF featuring the highest overall signal-to-noise. Some sciMMFs exhibit a very small signal-to-noise pentalty relative to the optimal case, whereas for others the penalty is extreme. For each sciMMF, the mean signal-to-noise relation is fit as a straight line, which is shown in green.}
\label{fig:snr_penalty}
\end{figure*}

In Section \ref{subsec:sciMMF} we showed that our sciMMF is very effective at suppressing the bias in the tSZ cluster observables caused by the cluster-correlated CIB emission. However, as anticipated in Section \ref{sec:theory}, foreground deprojection comes at the cost of a signal-to-noise penalty. Figure \ref{fig:snr_penalty} shows the signal-to-noise measurements for our fixed sciMMF catalogues plotted against those for our fixed random $100$--$857$\,GHz iMMF catalogue. The latter is the `optimal' catalogue, featuring the highest overall signal-to-noise, as it has both the largest number of frequency channels (six) and the smallest number of constraints (just one, that the tSZ signal is extracted). As expected, all our sciMMF catalogues suffer from a loss of signal-to-noise relative to the optimal case. This signal-to-noise penalty depends on the number of frequency channels that are taken into account and on the number and shapes of the SEDs that are deprojected. It ranges from very small in some instances (in particular, for $100$--$857$\,GHz with only deprojection of the mean CIB SED, but also for $100$--$857$\,GHz with additional deprojection of the first-order moment with respect to either $\beta_T$ or $\beta$, and for $100$--$545$\,GHz with deprojection of the mean SED), to very large in other cases. We recall that three of the sciMMFs retaining a significant signal-to-noise are virtually unbiased across signal-to-noise and redshift, as shown in \ref{subsec:cmmfbias}. We take them as our preferred sciMMFs and study their robustness against errors in the assumed deprojection SED in Section \ref{subsec:robustness}. These are the $100$--$854$\,GHz sciMMF deprojecting the CIB SED and its first-order moment with respect to $\beta$, the analogous sciMMF for $\beta_T$, and the $100$--$353$\,GHz sciMMF deprojecting only the mean CIB SED.

As the sciMMF signal-to-noise measurements are observed to be approximately proportional to the optimal iMMF ones, we fit the mean relation between them as a straight line with zero intersect using the standard least-squares method, fitting only for the slope. The fitted values of the slope, which should be interpreted as the overall signal-to-noise fraction relative to the optimal case, can be found in Table \ref{table}. We note that in order to mitigate against Malmquist bias, in the fit we only consider data points with $q_{\mathrm{iMMF}} > q_{\mathrm{min}}$, where $q_{\mathrm{min}}$ is chosen visually based on the minimum value of $q_{\mathrm{iMMF}}$ for which the sciMMF sample is seen to be approximately complete. We take $q_{\mathrm{min}} = 8$ for the $100$--$857$\,GHz mean, $100$--$857$\,GHz mean+$\beta$, $100$--$857$\,GHz mean+$\beta_T$, $100$--$545$\,GHz mean, and $100$--$353$\,GHz mean catalogues, $q_{\mathrm{min}} = 15$ for the $100$--$545$\,GHz mean+$\beta$ and $100$--$545$\,GHz mean+$\beta_T$ catalogues, and $q_{\mathrm{min}} = 20$ for the remaining catalogues.

\begin{table*}\label{table}
\centering
\begin{tabular}{l|lll}
  & \textbf{100--353\,GHz} & \textbf{100--545\,GHz} & \textbf{100--857\,GHz} \\
 \hline
\textbf{sciMMF mean} & $0.6516 \pm 0.0088$ & $ 0.918 \pm 0.010 $ & $ 0.964 \pm 0.011$   \\
\textbf{sciMMF mean+$\beta$} & $0.256 \pm 0.049 $ & $0.454 \pm 0.020 $ &  $0.875 \pm $ 0.013  \\
\textbf{sciMMF mean+$\beta_T$} & $0.0249 \pm 0.047 $ & $ 0.514 \pm 0.022 $ & $0.939 \pm 0.011 $ \\
\textbf{sciMMF mean+$\beta_T$+$\beta$} & $0.138 \pm 0.064$  & $0.261 \pm 0.049 $ & $0.375 \pm 0.036$ \\

\end{tabular}
\caption{Overall signal-to-noise fraction of our sciMMFs relative to the 100--857\,GHz iMMF. A higher value means a smaller signal-to-noise penalty. The quoted values correspond to the slopes of the lines fitted to the data in Figure \ref{fig:snr_penalty}.}
\end{table*}

\subsection{Robustness of results to the choice of deprojection SEDs}\label{subsec:robustness}

As demonstrated in Section \ref{subsec:cmmfbias}, our sciMMF is very effective at suppressing the CIB bias in the signal-to-noise, with several frequency and deprojection combinations being virtually unbiased. We recall that in our sciMMFs we used the same CIB SED that was used in the \Websky simulation, which is given by Eq. (\ref{eq:sed}), evaluated at the input parameter values (and at a fiducial redshift $z=0.2$). In particular, the dust emissivity index is $\beta = 1.6$ and the dust temperature at $z=0$, $T_{0} = 20.7$\,K. The first-order moments with respect to $\beta$ and $\beta_T = T_{0}^{-1}$ were similarly evaluated at the input parameter values. In short, apart from the redshift dependence, our deprojection SED perfectly matched the true cluster-correlated CIB SED of our mock data. 

A perfect match between the SED assumed in the sciMMF and the real SED of the cluster-correlated CIB will, however, never happen in practice, as the real CIB SED is determined with limited precision (see, e.g., \citealt{Cai2013,Planck2016XLIII,Melin2018}). In order to assess the impact of assuming a wrong deprojection SED, we produce and analyse a number of additional cluster catalogues assuming a `wrong' SED for our three preferred sciMMFs: (i) $100$--$857$\,GHz with deprojection of the mean CIB SED and its first-order moment with respect to $\beta$ (ii) same as in (i) but deprojecting the first-order moment with respect to $\beta_T$ instead, and (iii) $100$--$353$\,GHz with deprojection of the mean CIB SED. We also consider an additional sciMMF, $100$--$857$\,GHz with deprojection of only the mean CIB SED. For each of these four sciMMFs, we consider the CIB SED (and the first-order moments, where relevant) evaluated at (a) five values of $\beta$ linearly spaced between $\beta=1.4$ and $\beta=1.8$ and the true value of $T_0$, and (b) five values of $T_0$ linearly spaced between $16.8$\,K and $24.8$\,K and the true value of $\beta$. These two ranges of values encompass the values of $\beta$ and $T_0$ that are allowed observationally (e.g., \citealt{Planck2016XLIII}). We note that the third (i.e., central) value of both $\beta$ and $T_0$ that we consider corresponds to its true \Websky value. This leaves a total of 8 `wrong' cases for each sciMMF. We produce a cluster catalogue for each of them, with \texttt{SZiFi} applied in its fixed mode.

\begin{figure}
\centering
\includegraphics[width=0.4\textwidth]{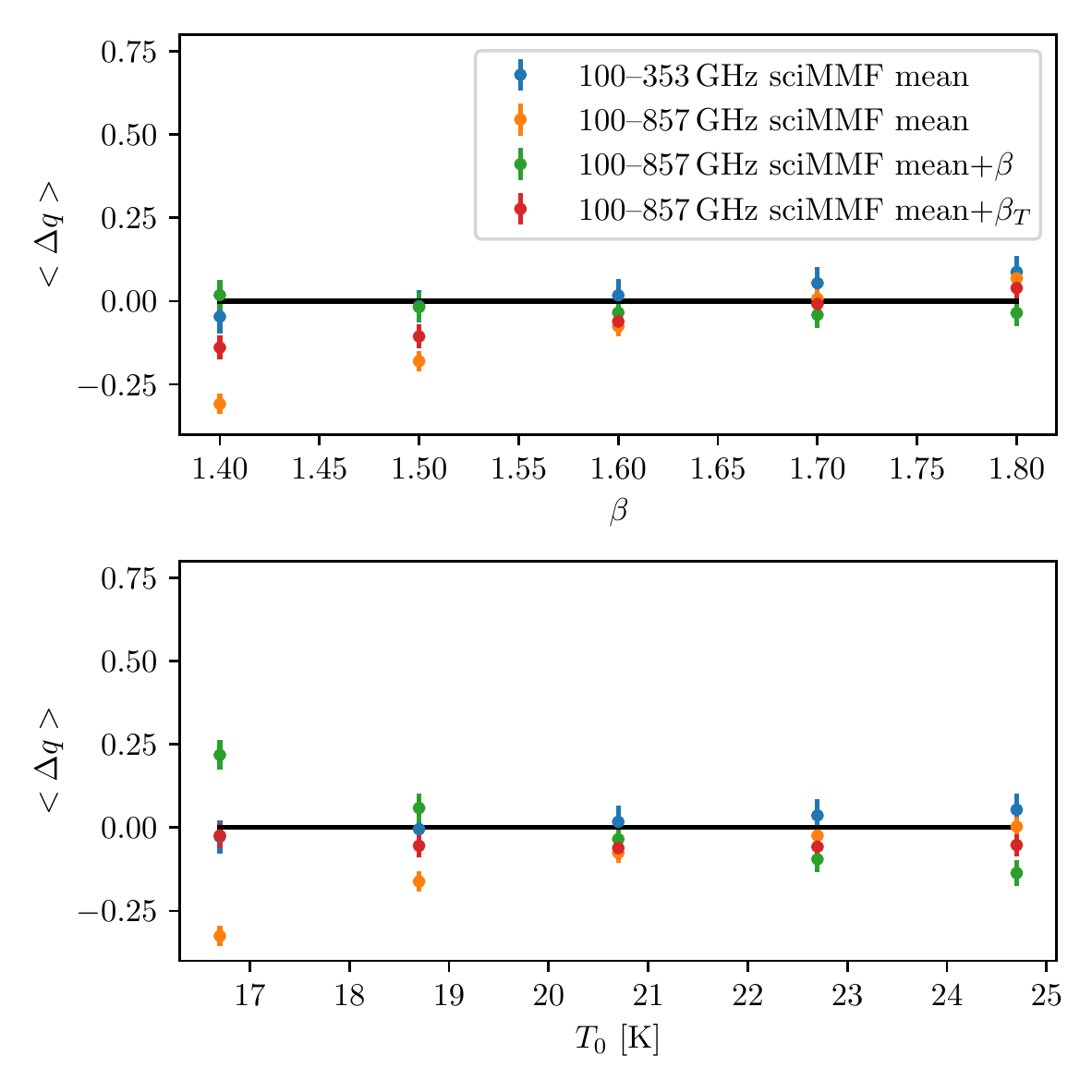}
\caption{Empirical mean of the signal-to-noise residuals for $\bar{q}_{\mathrm{t}} > 5$ for our four preferred sciMMF catalogues, shown as a function of the values of $\beta$ and $T_0$ at which the deprojection SED is evaluated. The central values of both panels correspond to the true values of $\beta$ and $T_0$ of the \Websky simulation.}
\label{fig:robust}
\end{figure}

The empirical mean of the signal-to-noise residuals of these cluster catalogues, calculated across the whole sample for $\bar{q}_{\mathrm{t}} > 5$, is shown in Figure \ref{fig:robust} as a function of $\beta$ and $T_0$. We note that the central data points in both panels are the same and correspond to the CIB SED evaluated at the true parameter values. The values of the central data points are $\left\langle \Delta q \right\rangle = -0.075 \pm 0.031$, $\left\langle \Delta q \right\rangle = -0.061 \pm 0.034$, $\left\langle \Delta q \right\rangle = -0.034 \pm 0.041 $, and $\left\langle \Delta q \right\rangle = 0.0174 \pm 0.048$ for the $100$--$857$\,GHz mean, mean+$\beta_T$, and mean+$\beta$ sciMMFs and the $100$--$353$\,GHz mean sciMMF, respectively. The two latter sciMMFs are thus completely unbiased on average (and, indeed, across signal-to-noise and redshift, see Section \ref{subsec:sciMMF}), whereas potentially a small negative bias is seen with $1.8$\,$\sigma$ and $2.4$\,$\sigma$ significance, respectively, for the two first sciMMFs. This bias is driven by the low-redshift clusters in the sample, as we discussed in Section \ref{subsec:cmmfbias}.

As either $\beta$ or $T_0$ are moved away from their true values, a systematic change in $\left\langle \Delta q \right\rangle$ is observed for the four sciMMFs. For the $100$--$353$\,GHz mean sciMMF (blue data points), the change is very small, with $\left\langle \Delta q \right\rangle$ remaining within $1$\,$\sigma$ of zero in most instances. For $100$--$857$\,GHz, the largest variation, and therefore bias, is observed when only the mean CIB SED is deprojected (orange data points). However, even in the most biased instances, with $\beta$ and $T_0$ take their minimum values, the bias is smaller than in the $100$--$857$\,GHz iMMF with no deprojection at all, for which $\left\langle \Delta q \right\rangle = -0.601 \pm 0.030$. Much smaller biases are observed, in general, if the first-order moments are also deprojected, demonstrating the power of the moment expansion approach of \citet{Chluba2017} to mitigate the impact of uncertainties in the assumed SED. In particular, as expected, when one parameter ($\beta$ or $T_0 = \beta_T^{-1}$) is varied, deprojecting the first-order moment with respect to it leads to a smaller change in $\left\langle \Delta q \right\rangle$ than if the first-order moment with respect to the other parameter is deprojected instead. %(Note that instead of computing the first-order moment with respect to $T_0$, we do so with respect to $\beta_T$, as this leads to better performance in general \citep{Chluba2017}, but a change in $T_0$ is obviously equivalent to a change in $\beta_T$.)

%Maybe add histograms?

\subsection{Adequacy of our sciMMF and comparison with a spectrally and spatially constrained MMF}\label{subsec:pointsource}

Our sciMMF has proven to be highly effective at eliminating the CIB-induced bias in the tSZ cluster observables. We recall that ours is a \emph{spectrally} constrained iMMF, in which one or several foregrounds with given SEDs are nulled, regardless of their spatial nature. A constrained matched filter in which a spatial template for the foreground(s) to be removed is assumed can also be constructed (see \citealt{Erler2019}). As shown in \citet{Erler2019}, this approach can be very effective at reducing contamination by point-like sources at the cluster centres. This spectrally \emph{and} spatially constrained matched filter could, in principle, also be used to tackle the CIB bias. 

\begin{figure}
\centering
\includegraphics[width=0.45\textwidth]{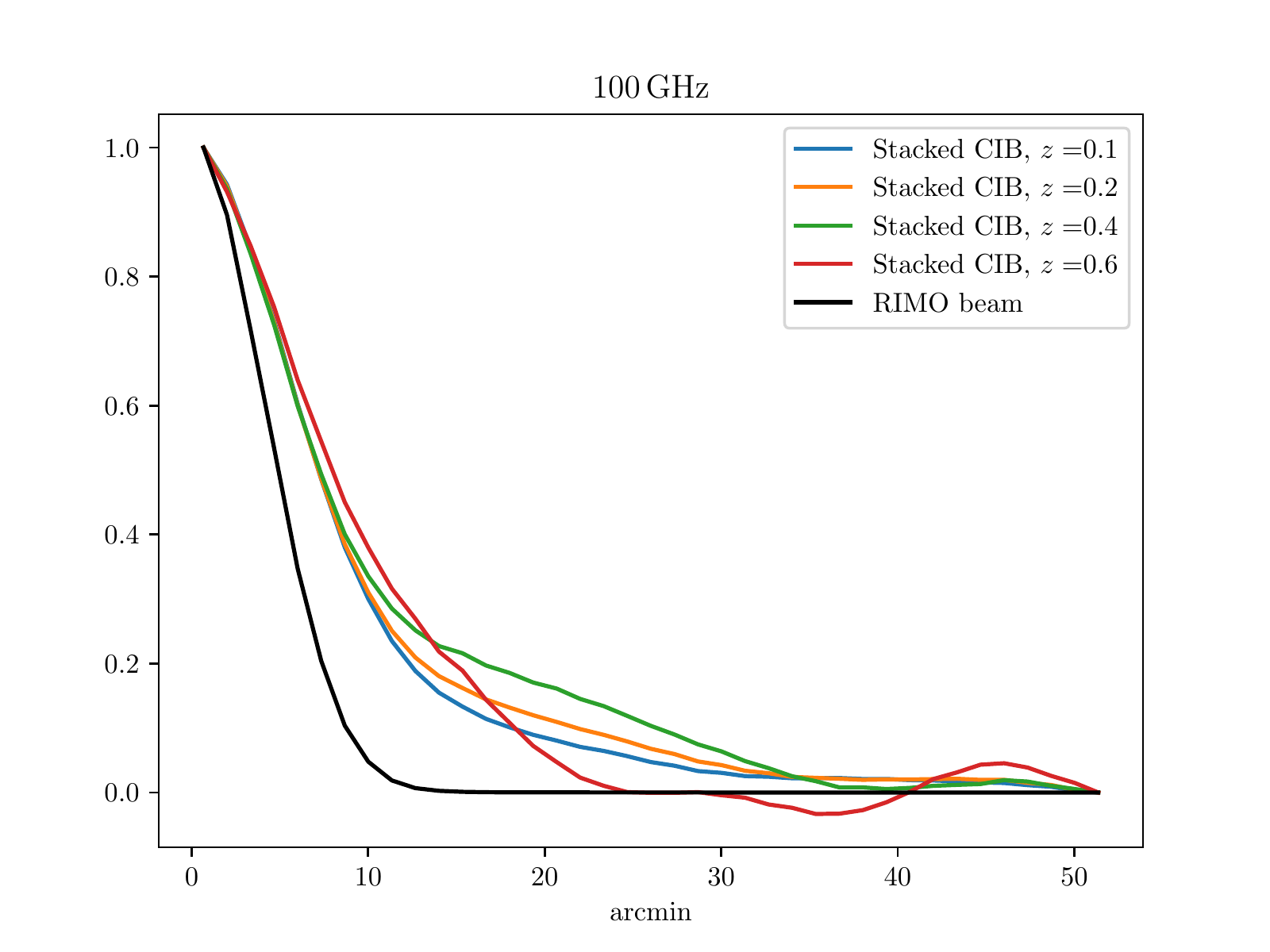}
\caption{\Websky CIB emission as seen by our \textit{Planck}-like experiment at 100\,GHz stacked at the location of the \Websky clusters with $M_{500} > 10^{14} M_{\odot}$ and redshifts within $z = 0.02$ of four reference redshifts. The \textit{Planck} 100\,GHz RIMO beam is also shown for comparison. We have subtracted the value of the stacked profiles at the maximum radius considered (51.36\,arcmin), effectively removing the uncorrelated CIB contribution, and we have rescaled the profiles so that they all have unit values at the centre. It is apparent that the cluster-correlated CIB emission is not point-like for an experiment with the resolution of \textit{Planck}. This is in agreement with observations \citep{Planck2016XXIII,Melin2018}.}
\label{fig:pointsource}
\end{figure}

However, the cluster-correlated CIB emission, which is what ought to be deprojected, is not a point-like source for an experiment with the resolution of \textit{Planck}. This can be seen in Figure \ref{fig:pointsource}, which shows the CIB emission from the \Websky simulation convolved by the \textit{Planck} beam at 100\,GHz stacked at the locations of the Wesbky clusters with $M_{500} > 10^{14} M_{\odot}$ and redshifts within $\Delta z = 0.02$ of four reference redshifts. We note that we have subtracted the value of the stacked profiles at the maximum radius considered (51.36\,arcmin), effectively removing the contribution from uncorrelated CIB emission, and we have rescaled the profiles so that they all have unit values at the centre. The \textit{Planck} isotropic beam at 100\,GHz is also shown for comparison. The cluster-correlated CIB emission is clearly not point-like for any of the redshifts considered, which encompass those of most of the \textit{Planck} clusters. This is also true for the remaining \textit{Planck} HFI channels, as they have smaller beams. These findings are consistent with observations, with extended dust emission at the locations of the \textit{Planck} clusters having been detected using \textit{Planck} data \citep{Planck2016XXIII,Melin2018}. 

If the constrained matched filters of \citet{Erler2019} were to be used, a model for the cluster-correlated CIB emission (effectively, a model for the tSZ--CIB correlation; see, e.g., \citealt{Addison2012,Maniyar2021}) would be needed, which would add additional modelling uncertainties. Since our sciMMF only requires knowledge of the CIB SED, and since the impact of our imperfect knowledge of it can be very effectively mitigated using the moment-expansion approach, with an acceptable associated signal-to-noise penalty, we favour of sciMMF as a tool to tackle the CIB bias. A detailed comparison between our sciMMF and the constrained matched filters of \citet{Erler2019}, however, remains an interesting possibility to be considered in future work. %add references about how well this is measured

%mention some references / detections??

\section{Signal-to-noise penalty forecasts for ACT, SPT, and SO}\label{subsec:forecasts}

In this work we have considered our sciMMF in the context of \textit{Planck}, showing that it can be highly effective at removing the CIB-induced bias from the cluster tSZ observables while incurring an acceptable signal-to-noise penalty. A full analysis of the performance of our sciMMF for other experiments is beyond the scope of this paper. However, in this section we estimate the mean signal-to-noise loss due to CIB deprojection for three additional CMB experiments: the Atacama Cosmology Telescope with the Advanced ACTPol receiver (ACT; \citealt{Henderson2016}), the South Pole Telescope with the SPTpol instrument (SPTpol; \citealt{Austermann2012}), and the Large Aperture Telescope (LAT) of the upcoming Simons Observatory (SO; \citealt{SO2019}), both with its baseline and goal specifications. The exact observational specifications that we assume for each experiment are detailed in Table~\ref{table_exp22}. In particular, those of ACT are intended to be representative of the real observational specifications of the latest ACT cluster survey \citep{Hilton2020}, and those of SPTpol of the latest SPT cluster survey \citep{Bleem2020}.

In order to estimate the mean loss of signal-to-noise due to CIB deprojection, we generate mock sky maps for each experiment in the same way that we did for \textit{Planck} (see Section~\ref{sec:sims}). Namely, we convolve the \Websky sky at each frequency band with the instrument beam, using Gaussian beams, and add white noise with the appropriate noise levels (see Table \ref{table_exp22}). As with our \textit{Planck}-like mock observations, the maps contain the following components: tSZ, kSZ, CIB, CMB, and noise. For each experiment, we run \texttt{SZiFi} on ten sky cut-outs, defined in the same way as in our analysis of \textit{Planck}-like data, in order to estimate the noise covariance in the maps, and compute the average noise covariance across the ten cut-outs. With this covariance in hand, for each experiment we calculate the signal-to-noise of a reference cluster of $M_{500}= 5 \times 10^{14} M_{\odot}$ and $z = 0.2$ assuming the \citet{Arnaud2010} pressure profile. We do so for the standard iMMF and for our sciMMF, deprojecting the \Websky CIB SED evaluated at the true parameter values and $z=0.2$. For SO, we also consider the deprojection of the first-order moments with respect to $\beta$ and $\beta_T$, individually and together. Note that this is not possible for ACT and SPTpol, as they only have two frequency channels, as opposed to the six channels of SO. We then compute the ratio between the signal-to-noise for the different sciMMFs and that for the iMMF. As we saw in Section~\ref{subsec:snrpenalty}, this ratio is approximately constant regardless of the value of the iMMF signal-to-noise (i.e., of the cluster mass and redshift), so our calculation for a cluster with given mass and redshift can be interpreted as an estimate of the overall signal-to-noise of each sciMMF relative to the iMMF.

\begin{figure}
\centering
\includegraphics[width=0.45\textwidth,trim={00mm 0mm 0mm 10mm},clip]{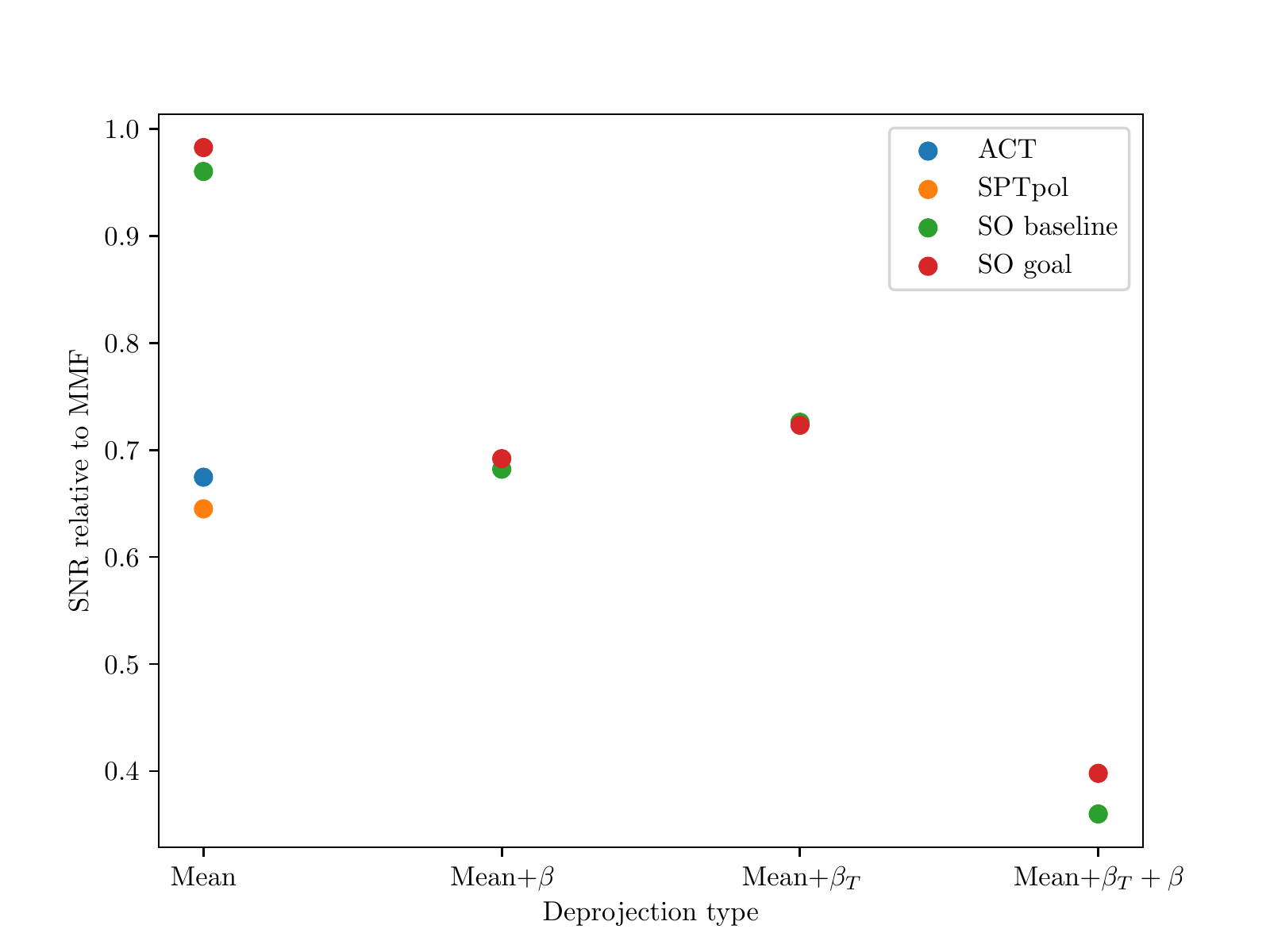}
\caption{Forecasts of the mean signal-to-noise of the sciMMF relative to that of the iMMF for ACT, SPTpol, and SO. Several deprojections are considered (mean CIB SED and first-order moments). A larger value corresponds to a smaller signal-to-noise penalty due to foreground removal.}
\label{fig:snr_forecast}
\end{figure}

These sciMMF to iMMF signal-to-noise ratios are shown in Figure \ref{fig:snr_forecast}. As in Table \ref{table}, a larger ratio means a smaller signal-to-noise penalty. For SO, both in its baseline and goal configurations, deprojecting only the mean CIB SED results in a very small (few percent) loss of signal-to-noise, as it was the case for \textit{Planck}. Further deprojection of one first-order moments increases the signal-to-noise penalty to about $30$\,\%. This is higher than what was observed for \textit{Planck}, which was at the level of 5--10\,\%, which we attribute to the lack of high-frequency coverage in SO, where the CIB SED is less degenerate with the tSZ one. However, it can still be acceptable, particularly given the expected size of the SO cluster sample (about 20\,000 objects, \citealt{SO2019}). Deprojecting both first-order moments at the same time, however, leads to a much higher signal-to-noise penalty (about 60\,\%). For ACT and SPTpol, as we have already noted, we can only deproject one single SED. Deprojecting the mean CIB SED leads to a $\sim 25$\,\% signal-to-noise penalty for both experiments. As for SO, although this penalty may seem quite significant, CIB deprojection would be an interesting possibility to consider in both experiments, given the statistical power of their cluster samples.

\begin{table*}
\centering
\begin{tabular}{llll}
\hline
\textbf{Experiment} & \textbf{Frequencies} & \textbf{FWHM of beams} & \textbf{Noise levels}  \\
\hline
 ACT & 100, 145\,GHz & 2.2, 1.4\,arcmin  &  33, 27\,$\mu$K\,arcmin  \\
SPTpol & 93, 145\,GHz & 1.7, 1.2\,arcmin  & 60, 33 \,$\mu$K\,arcmin    \\
SO baseline & 27, 39, 93, 145, 225, 278\,GHz & 7.4, 5.1, 2.2, 1.4, 1.0, 0.9\,arcmin  & 71, 36, 8, 10, 22, 54\,$\mu$K\,arcmin   \\
SO goal & 27, 39, 93, 145, 225, 278\,GHz & 7.4, 5.1, 2.2, 1.4, 1.0, 0.9\,arcmin  & 52, 27, 5.8, 6.3, 15, 37\,$\mu$K\,arcmin.

\end{tabular}
\caption{Experimental specifications assumed in the computation of signal-to-noise penalty forecasts for ACT, SPTpol, and the Simons Observatory (see Section \ref{subsec:forecasts}).}
\label{table_exp22}
\end{table*}

\section{Conclusion}\label{sec:conclusion}

In this paper we have studied the impact of the cluster-correlated CIB emission on galaxy cluster detection with multi-frequency matched filters (MMFs) in the context of \textit{Planck}. We have done so by applying the iterative MMF of \citet{Zubeldia2022}, as implemented in \texttt{SZiFi}, to mock \textit{Planck}-like data produced with the sky maps of the state-of-the-art \Websky simulation. We have considered three frequency combinations (100--353\,GHz, 100-545\,GHz, and 100--857\,GHz) and have found a significant bias in the signal-to-noise of the detections for the three of them, with a mean bias of about $0.5\,\sigma$ \emph{per cluster}. We have also found a 5--10\,\% bias in the matched-filter estimates of the cluster central Compton-$y$ parameter, $y_0$. We have shown that the bias in the signal-to-noise translates into a bias in the survey completeness function and, as a result, in the observed cluster number counts. In particular, if all the \textit{Planck} HFI channels are used, as it was the case in \citet{Planck2016xxvii}, about $20$\,\% fewer objects are detected with respect to the case in which the tSZ--CIB correlation is removed by spatially randomising the simulated CIB. Unless properly accounted for, this difference in the number counts could potentially bias derived cosmological constraints, a possibility that we will investigate in future work. We note that the CIB-induced biases that we have observed, both in the number counts and in the Compton-$y$ amplitudes, are larger than those seen in \citet{Melin2018}. We attribute this discrepancy to the difference in the dust emission models assumed in the production of the mock data analysed in each study. %As we have pointed out, the Websky CIB power spectra and CIB--tSZ cross-correlation are in full agreement with \textit{Planck} data, but 

We have then introduced a spectrally constrained iterative MMF, or sciMMF, which is designed to remove the signals from a set of foregrounds with given SEDs, and we have applied it to our simulated \textit{Planck}-like data. We have shown that our sciMMF can be highly effective at suppressing the CIB-induced bias from the cluster observables (signal-to-noise and $y_0$) and from the survey completeness while incurring only a small signal-to-noise penalty (see Figures \ref{fig:dep_q}, \ref{fig:dep_com}, \ref{fig:snr_penalty}, and \ref{fig:dep_y0}). In particular, we have identified three cases featuring an acceptable loss of signal-to-noise for which the cluster observables are virtually unbiased across signal-to-noise and redshift. These are the 100--857\,GHz sciMMF with deprojection of the mean CIB SED and its first-order moment with respect to either the dust emissivity index $\beta$ or the dust temperature parameter $\beta_T = T_0^{-1}$ (87.5\,\% and 93.9\,\% of signal-to-noise retained relative to the optimal unconstrained iMMF, respectively), and the 100--353\,GHz sciMMF with deprojection of the mean CIB SED (65.16\,\% signal-to-noise retained). 

We have quantified the robustness of the bias suppression provided by these sciMMFs to errors in the SED assumed for deprojection, remarking that the deprojection SED is the only modelling assumption made about the foreground to be deprojected. We have found that our preferred sciMMFs remain unbiased or with a very small bias if $\beta$ or $\beta_T$ are varied within a reasonable range allowed by observations (see Figure~\ref{fig:robust}). This is a very promising result towards the application of our sciMMF to real observations, for which the exact SED of the cluster-correlated CIB is not perfectly determined.

Finally, we have produced estimates of the sciMMF signal-to-noise loss for ACT, SPTpol and SO, showing that CIB deprojection is achievable with a small-to-moderate signal-to-noise penalty for all of them. Given the statistical power of the cluster samples constructed with data from these experiments, CIB deprojection through our sciMMF thus constitutes an interesting possibility to be considered for all of them. This is particularly true for SO, which is forecast to detect about $20,000$ clusters and to probe the low-mass cluster regime, where CIB contamination is expected to be more significant \citep{Melin2018}.

%We note that the CIB-induced biases that we have observed, both in the number counts and in the Compton-$y$ amplitudes, are larger than those seen in \citet{Melin2018}. We have attributed this discrepancy to the difference in the dust emission models assumed in the production of the mock data analysed in each study. As we have pointed out, the Websky CIB power spectra and CIB--tSZ cross-correlation are in full agreement with \textit{Planck} data, but

\section*{Acknowledgements}

The authors would like to thank Jean-Baptiste Melin for useful comments on the manuscipt.

This work was supported by the ERC Consolidator Grant {\it CMBSPEC} (No.~725456) as part of the European Union's Horizon 2020 research and innovation program. IZ was also supported by an STFC Consolidated Grant, and JC by the Royal Society as a Royal Society URF at the University of Manchester

%%%%%%%%%%%%%%%%%%%%%%%%%%%%%%%%%%%%%%%%%%%%%%%%%%
\section*{Data Availability}
 
The data underlying this article will be shared on reasonable request to the corresponding author.

%%%%%%%%%%%%%%%%%%%% REFERENCES %%%%%%%%%%%%%%%%%%

% The best way to enter references is to use BibTeX:

\bibliographystyle{mnras}
\bibliography{references} % if your bibtex file is called example.bib

% Alternatively you could enter them by hand, like this:
% This method is tedious and prone to error if you have lots of references
%\begin{thebibliography}{99}
%\bibitem[\protect\citeauthoryear{Author}{2012}]{Author2012}
%Author A.~N., 2013, Journal of Improbable Astronomy, 1, 1
%\bibitem[\protect\citeauthoryear{Others}{2013}]{Others2013}
%Others S., 2012, Journal of Interesting Stuff, 17, 198
%\end{thebibliography}

%%%%%%%%%%%%%%%%%%%%%%%%%%%%%%%%%%%%%%%%%%%%%%%%%%

%%%%%%%%%%%%%%%%% APPENDICES %%%%%%%%%%%%%%%%%%%%%

\appendix

\section{Results for \lowercase{$y_0$}}\label{appendix:y0}

In this paper we have discussed the cluster CIB bias and its suppression by our sciMMF in terms of our preferred cluster observable, the signal-to-noise $q$. An alternative observable is the matched filter amplitude parameter, $\hat{y}_0$, which is an estimator of the cluster Compton-$y$ parameter at the cluster's centre, $y_0$ (see Section \ref{sec:theory}). In this appendix we consider this observable, showing analogous results to those presented in Section \ref{sec:results} for $q$. We use the same \textit{Planck}-like mock catalogues that were analysed in Section \ref{sec:results}, which are described in Section \ref{subsec:catalogues}.

\begin{figure*}
\centering
\includegraphics[width=0.7\textwidth,trim={00mm 5mm 0mm 10mm},clip]{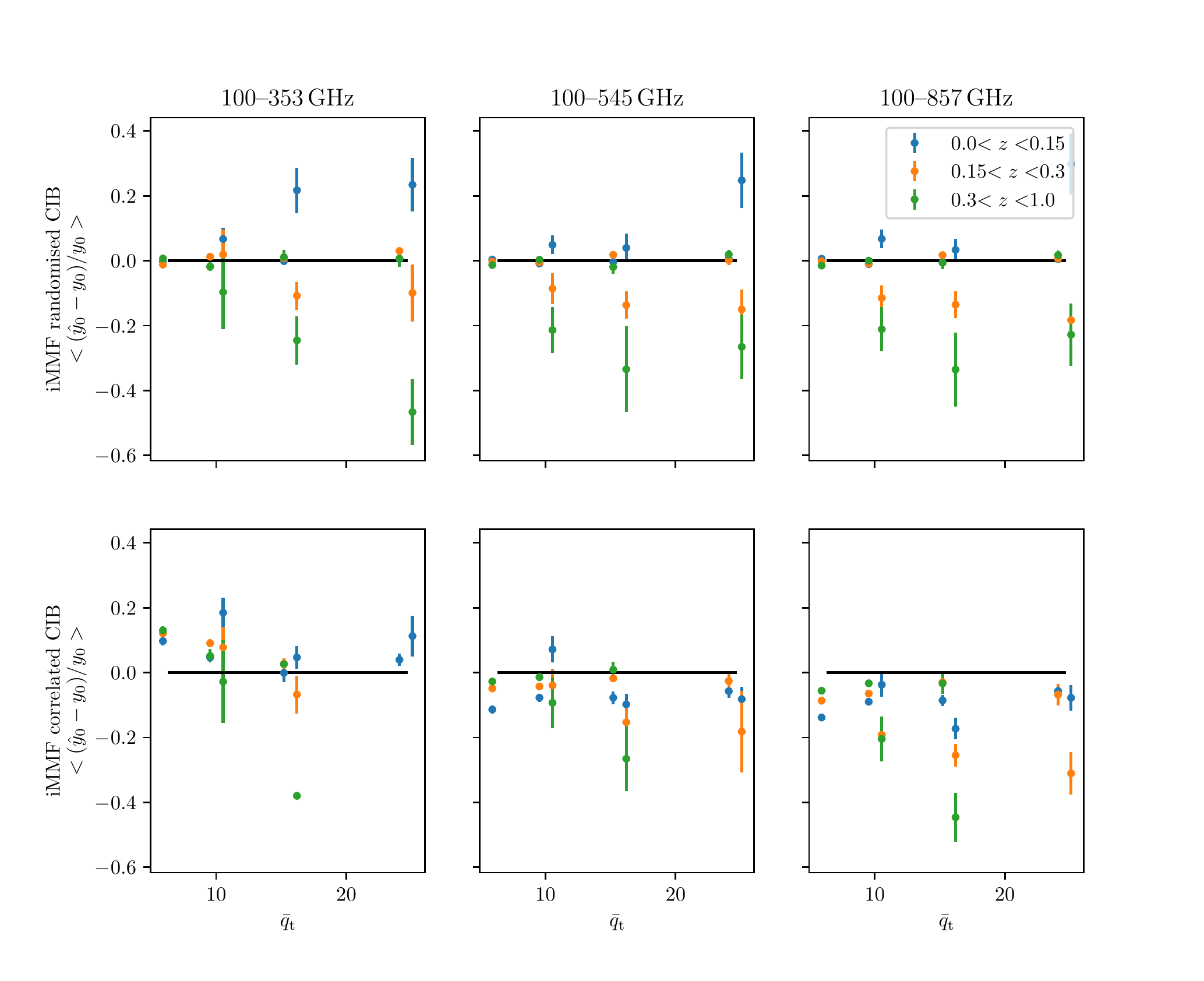}
\caption{Mean of the $\hat{y}_0$ residuals for our iMMF catalogues, both if the CIB is spatially correlated with the tSZ signal (as it would in a real experiment; lower panels) and if it is randomised (upper panels). A significant frequency and redshift-dependent bias is seen in the former case, whereas no bias is seen in the latter. The data points on the left-hand side of each signal-to-noise bin correspond to the catalogues obtained with \texttt{SZiFi} run in its fixed mode, with the signal-to-noise extracted at the clusters' true parameters (angular and sky location), whereas those on the right-hand side of each bin correspond to the blind cluster catalogues obtained with \texttt{SZiFi} in its cluster-finding mode. Figure \ref{fig:corrvsuncorr_q} is an analogous plot for the signal-to-noise.}
\label{fig:corrvsuncorr_y0_find}
\end{figure*}

As we discussed in Section \ref{sec:results}, the tSZ--CIB spatial correlation leads to a bias in the iMMF signal-to-noise (see Figure \ref{fig:corrvsuncorr_q}). A bias is also observed for $\hat{y}_0$. This can be seen in Figure \ref{fig:corrvsuncorr_y0_find}, a plot analogous to Figure \ref{fig:corrvsuncorr_q} showing the empirical mean of the $\hat{y_0}$ residuals, $(\hat{y_0}-y_0)/y_0$, for our iMMF catalogues if the CIB is properly correlated with the tSZ field (lower panels), and if it has been randomised (upper panels). For each signal-to-noise bin, the data points on the left-hand side correspond to the fixed case, in which $\hat{y}_0$ is extracted at the cluster true parameter values (angular size and sky location), whereas those on the right-hand side correspond to the cluster-finding case, in which the clusters are blindly detected as peaks in the signal-to-noise distribution. In the latter case, $\hat{y_0}$ is given by the MMF evaluated at the optimal cluster parameter values, i.e., those for which the signal-to-noise is maximised.

It is apparent that in the absence of tSZ--CIB correlation, $\hat{y}_0$ is an unbiased estimator in the fixed case, whereas the tSZ--CIB correlation induces a $5$--$10$\,\% bias. The behaviour of the bias as a function of true signal-to-noise and redshift is very similar to that of the CIB bias in the signal-to-noise observed in Figure \ref{fig:corrvsuncorr_q}. We note that the bias is larger than the 1--2\,\% bias in the integrated MMF cluster Compton-$y$, $Y_{500}$, reported in \citet{Melin2018} (see their Figure 6). We attribute this difference to the choice of dust emission models used to produce the mock data analysed in each study (see the discussion in Section \ref{subsec:number_counts}).

In the cluster-finding case, a larger bias is observed even in the uncorrelated case (upper panels). This is due to the fact that $\hat{y}_{0}$ is not evaluated at the true values of the cluster parameters (angular size and sky location), but at those maximising the signal-to-noise. As discussed in \citet{Zubeldia2021}, these optimal parameter values can be thought of as maximum-likelihood estimates and are, as a consequence, biased in general, thus leading to a biased $\hat{y}_{0}$. This optimisation bias is the main reason why we choose the signal-to-noise as our preferred observable, as for it the effect of optimisation can be accounted for very easily using the simple prescription of \citet{Zubeldia2021}.

\begin{figure*}
\centering
\includegraphics[width=0.8\textwidth,trim={00mm 15mm 0mm 20mm},clip]{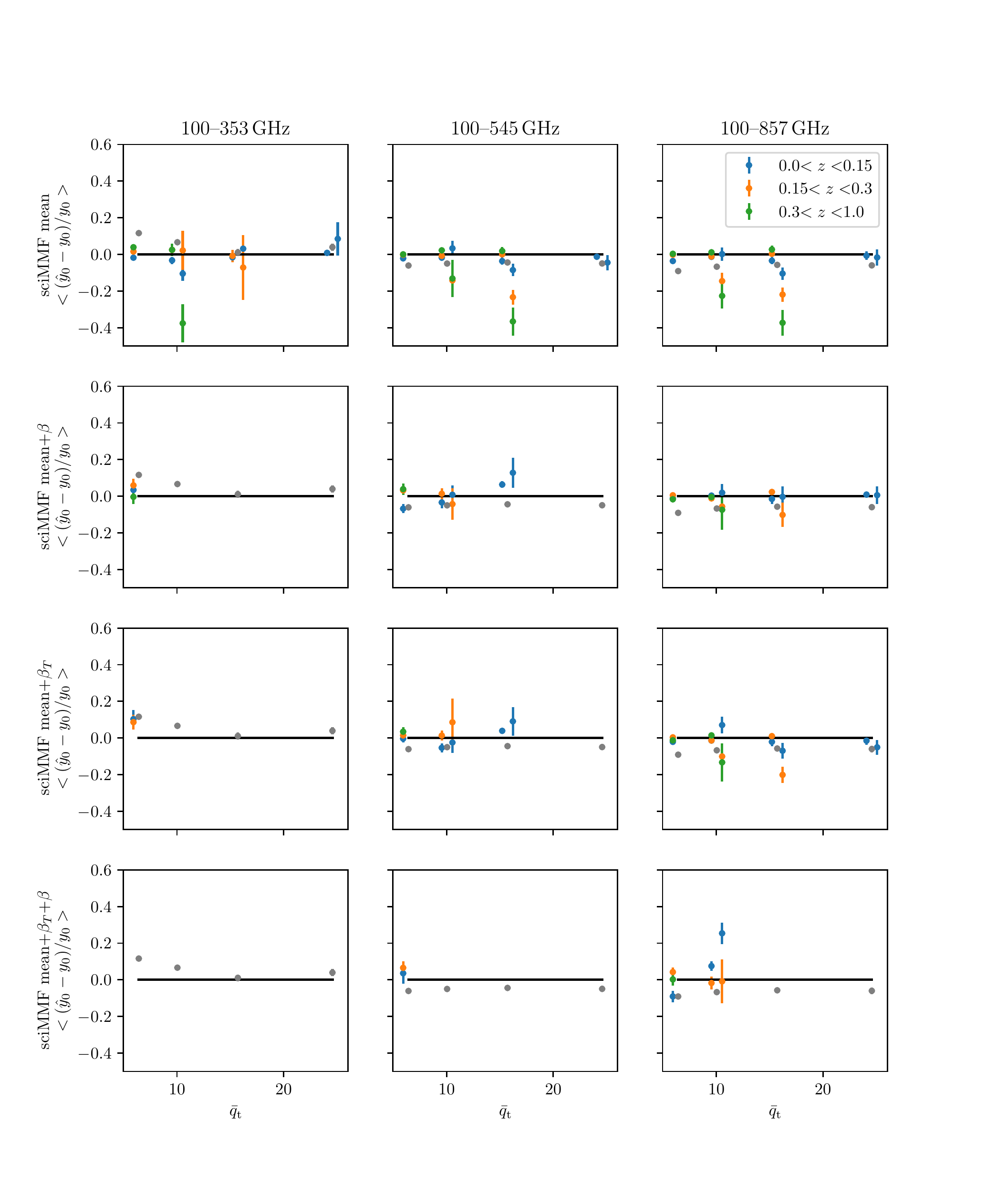}
\caption{Empirical mean of the $\hat{y}_0$ residuals binned in true signal-to-noise $\bar{q}_{\mathrm{t}}$ and redshift for our simulated sciMMF cluster catalogues. For comparison, the mean of the $\hat{y}_0$ residuals for the iMMF cluster catalogues, averaged in a single redshift bin, are also shown (grey data points). Figure \ref{fig:dep_q} is an analogous plot for the signal-to-noise.}
\label{fig:dep_y0}
\end{figure*}

As was the case for the signal-to-noise, CIB deprojection is also very effective at suppressing the CIB bias from $\hat{y}_0$. Figure \ref{fig:dep_y0} shows the $\hat{y_0}$ residuals for our set of sciMMFs, for both the fixed and the cluster-finding cases (data points on the left-hand and right-hand side of each signal-to-noise bin, respectively). In the fixed case, the situation is very similar to that seen for the signal-to-noise in Figure \ref{fig:dep_q}, with the 100--353\,GHz sciMMF with deprojection of the mean CIB SED and the 100--857\,GHz sciMMFs with deprojection of the mean CIB SED and its first-order moment with respect to either $\beta$ or $\beta_T$ providing a virtually unbiased estimator while retaining an acceptable signal-to-noise penalty. In the cluster-finding case, the optimisation bias is also present, significantly biasing some instances (e.g., the 100--857\,GHz sciMMF with deprojection of the mean CIB SED and its first-order moment with respect to either $\beta_T$).% We remark that  that this was not the case for the signal-to-noise (see Figure \ref{fig:dep_q}), for which the optimisation bias can be accurately accounted for.

\section{Cluster confusion and its impact on the completeness}\label{appendix:confusion}

As discussed in Section \ref{subsec:completeness}, when \texttt{SZiFi} is run in its cluster-finding mode, with clusters detected blindly, and the CIB is randomised, the empirical survey completeness function is about 5\,\% lower than expected. Thus, although the signal-to-noise measurements are unbiased (see Figure \ref{fig:corrvsuncorr_q}), as is the completeness in the fixed mode, some clusters are missing. The cluster-finding completeness is also observed to be low with respect to its theoretical expectation for several of our sciMMFs for which the signal-to-noise and the completeness in the fixed case are otherwise unbiased (see Figure \ref{fig:dep_com}).

\begin{figure}
\centering
\includegraphics[width=0.4\textwidth,trim={00mm 0mm 0mm 10mm},clip]{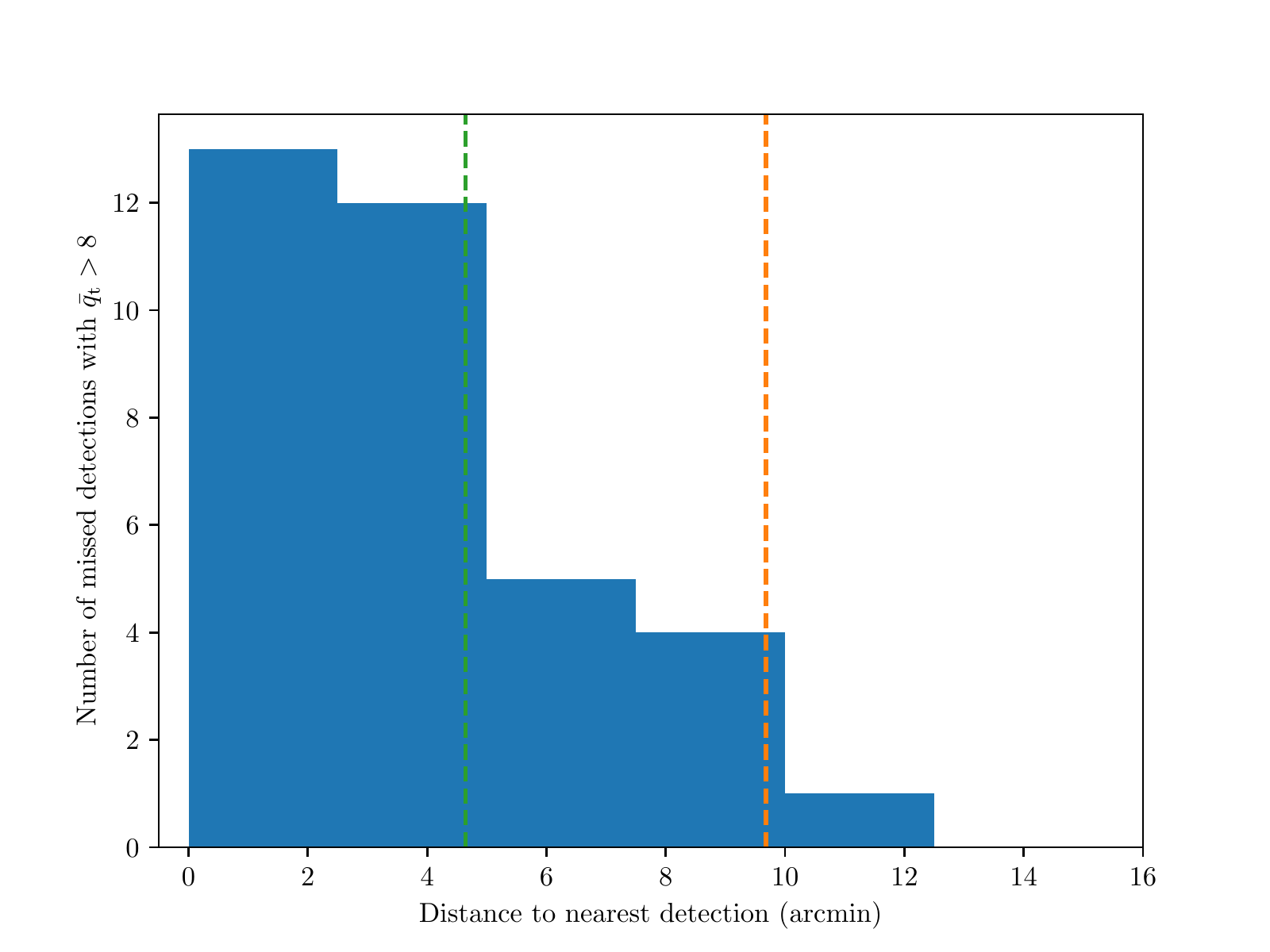}
\caption{Histogram of the distance to the closest detection for the undetected clusters with true signal-to-noise $\bar{q}_{\mathrm{t}} > 8$ in our randomised 100--857\,GHz iMMF catalogue. The FWHM of the \textit{Planck} 100\,GHz and 857\,GHz channels are also shown for reference (orange and green dashed vertical lines, respectively).}
\label{fig:missed}
\end{figure}

These missing clusters are clusters that are too close in the sky to another cluster to be detected individually. Instead, \texttt{SZiFi} identifies the joint signal as a single detection, and therefore one, or several, of them appear as missing when the \Websky input catalogue is compared with the blindly-detected observed catalogue. Figure \ref{fig:missed} shows a histogram of the distances between the undetected clusters with true signal-to-noise $\bar{q}_{\mathrm{t}} > 8$ and the closest detection to them for our randomised 100--857\,GHz iMMF catalogue. The FWHMs of the lowest and highest-frequency \textit{Planck} HFI channels are also shown for comparison (orange and green lines, respectively). Since the signal-to-noise has unit standard deviation (to very high accuracy in this case, approximately in others, as discussed in Section \ref{subsec:std}) and the signal-to-noise threshold used in the construction of the blind catalogue is $q=5$, all clusters with true signal-to-noise $\bar{q}_{\mathrm{t}} > 8$ should have been detected with very high probability. Indeed, in the analogous fixed catalogue, if the same threshold is imposed on the observed signal-to-noise, all the clusters with $\bar{q}_{\mathrm{t}} > 8$ are included in the sample. However, since these clusters are within a distance to another cluster that is comparable to the instrument beams, they are not blindly detected.

\begin{figure}
\centering
\includegraphics[width=0.4\textwidth]{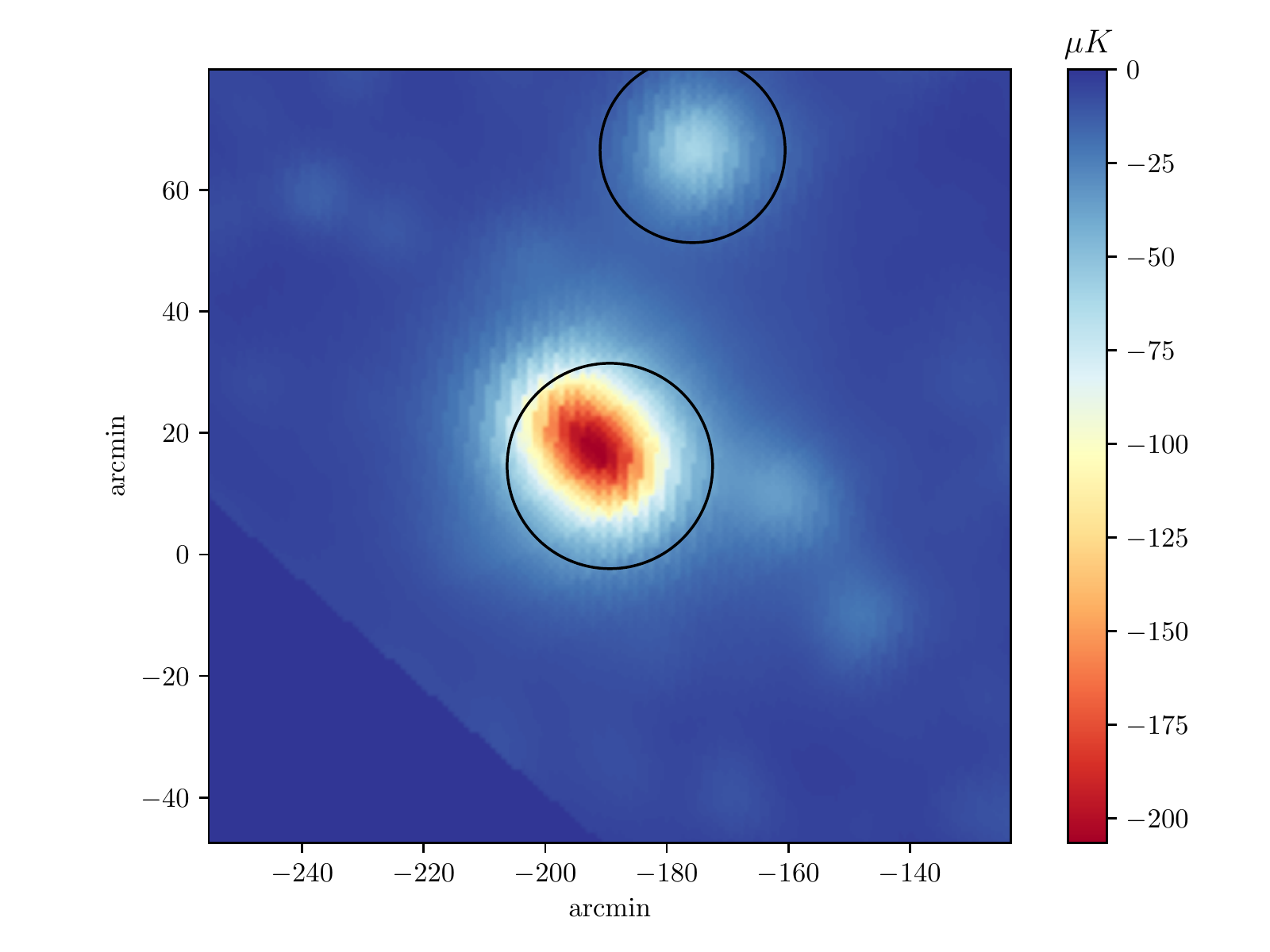}
\caption{\Websky tSZ signal at 100\,GHz for a small cut-out from one of our sky tiles, convolved by the \textit{Planck} beam. Blind detections in our randomised 100--857\,GHz iMMF catalogue are shown as black open circles. The central signal corresponds in fact to two halos of comparable masses, which are detected as a single object.}
\label{fig:missed_map}
\end{figure}

As an illustration of this `confusion' effect, Figure \ref{fig:missed_map} shows the beam-convolved \Websky tSZ signal at 100\,GHz for a small cut-out from one of our sky tiles. The beam used for convolution is the \textit{Planck} 100\,GHz isotropic beam, and the blind detections with signal-to-noise greater than 5 in the randomised 100--857\,GHz iMMF catalogue are shown as black open circles. The detected object at the centre of the cut-out, which has a markedly non-spherical shape (note that the \Websky clusters are spherical), corresponds to two clusters of comparable masses ($M_{500} = 6.93 \times 10^{14} M_{\odot}$ and $M_{500} = 6.01 \times 10^{14} M_{\odot}$), which cannot be distinguished as individual objects by \texttt{SZiFi}.

The bias in the completeness caused by this confusion effect could be accounted for in a likelihood analysis at the theory level, e.g., by modifying the theoretical completeness of Eq. (\ref{erf}) with a `confusion' factor quantifying the fraction of clusters that are missed as a function, e.g., of the true signal-to-noise. This factor could, in turn, be calibrated against simulated data. Alternatively, additional data (e.g., from optical and/or X-ray observations) could be taken into account in order to assess whether the blind detections correspond to one or several clusters. We note that a detailed investigation of this effect and how account for it is beyond the scope of this paper.

%This `confusion' effect is illustrated further in Figure \ref{fig:completeness_confusion}, which depicts the empirical completeness for our three randomised iMMF catalogues (blue data points), along with its theoretical expectation, given by Eq. (\ref{erf}) with the optimisation bias correction (orange curve). In addition,  with the empirical completeness for the same catalogues if the input catalogue against which the observed one is matched is pre-processed with a friends-of-friends that merges clusters within $10$\,arcmin of each other. We note that the same algorithm is applied to our observed catalogues, which, as discussed in \citet{Zubeldia2022}, efficiently merges raw detections associated to single clusters. 

%\begin{figure*}
%\centering
%\includegraphics[width=0.7\textwidth]{results_completeness_merged.pdf}
%\caption{Completeness.}
%\label{fig:completeness_confusion}
%\end{figure*}

%%%%%%%%%%%%%%%%%%%%%%%%%%%%%%%%%%%%%%%%%%%%%%%%%%

% Don't change these lines
\bsp	% typesetting comment
\label{lastpage}
\end{document}